\documentclass[aps,pra,reprint,
							floatfix,
						 	twocolumn,
							superscriptaddress,
				            notitlepage,
							footinbib,
							floatfix,
						longbibliography]{revtex4-1}

\usepackage{graphicx}
\usepackage{dcolumn}
\usepackage{bm}		
\usepackage{amsmath}
\usepackage{amssymb}
\usepackage[utf8]{inputenc}

\usepackage[normalem]{ulem}

\usepackage[usenames,dvipsnames]{xcolor}
\usepackage{slashed}
\usepackage{booktabs}					
\usepackage{microtype}

\usepackage{rotating}

\DeclareSymbolFont{rsfs}{U}{rsfs}{m}{n}
\DeclareSymbolFontAlphabet{\mathrsfs}{rsfs}
\usepackage[mathscr]{eucal}	

\usepackage{natbib}

\usepackage{hyperref}
\hypersetup{colorlinks=true,citecolor=MidnightBlue,urlcolor=BrickRed,linkcolor=Green}

\renewcommand{\d}{\mathrm d}
\let\ud\d

\newcommand{\tr}{{\rm tr}\,}
\newcommand{\TR}[1]{\, \tr \! \left[#1\right] }
\newcommand{\Ai}{\mathrm{Ai}\,}

\let\vec\boldsymbol
\let\up\uparrow
\let\down\downarrow

\newcommand{\UP}{\mathsf{UP}}
\newcommand{\IP}{\mathsf{IP}}
\newcommand{\FP}{\mathsf{FP}}
\newcommand{\PC}{\mathsf{PC}}
\newcommand{\EP}{\mathsf{EP}}
\newcommand{\PP}{\mathsf{PP}}

\newcommand{\shd}{{\mathrm{sign}\,(\dot h) }}
\newcommand{\shdp}{{\mathrm{sign}\,(\dot h (\varphi)) } }

\newcommand{\eqnref}[1]{Eq.~\eqref{#1}}

\newcommand{\be}{\begin{equation}}
\newcommand{\ee}{\end{equation}}
\newcommand{\bi}{\begin{itemize}}
\newcommand{\ei}{\end{itemize}}
\newcommand{\bea}{\begin{eqnarray}}
\newcommand{\eea}{\end{eqnarray}}

\newcommand{\eps}{\varepsilon}

\newcommand{\figref}[1]{Fig. \ref{#1}}

\newcommand{\chip}{\chi_p}
\newcommand{\chik}{\chi_k}

\begin{document}

\title{Spin and polarisation dependent LCFA rates for nonlinear Compton and Breit-Wheeler processes}

\author{D. Seipt}
\email{d.seipt@hi-jena.gsi.de}
\affiliation{Helmholtz Institut Jena, Fr\"obelstieg~3, 07743 Jena, Germany}
\affiliation{GSI Helmholtzzentrum für Schwerionenforschung GmbH, Planckstrasse 1, 64291 Darmstadt, Germany}
\affiliation{The G\'erard Mourou Center for Ultrafast Optical Science, University of Michigan, Ann Arbor, Michigan 48109, USA}

\author{B. King}
\email{b.king@plymouth.ac.uk}
\affiliation{Centre for Mathematical Sciences, University of Plymouth, Plymouth, PL4 8AA, United Kingdom}

\begin{abstract}
In this paper we derive and discuss the completely spin- and photon-polarisation dependent probability rates for nonlinear Compton scattering and nonlinear Breit-Wheeler pair production.
The locally constant field approximation, which is essential for applications in plasma--QED simulation codes, is rigorously derived from the strong-field QED matrix elements in the Furry picture for a general plane-wave background field.
We discuss important polarisation correlation effects in the spectra of both processes.
Asymptotic limits for both small and large values of $\chi$ are derived and their spin/polarisation dependence is discussed.
\end{abstract}


    \maketitle

    \section{Introduction}

    High-intensity laser experiments have now reached the point of being able to investigate the strong-field regime of quantum electrodynamics (QED). In this novel regime, elementary particles, such as electrons and photons, interact nonperturbatively with extremely strong electromagnetic fields. Recent measurements performed at the RAL-CLF's Gemini laser already hint at the relevance of quantum effects in radiation reaction \cite{cole18,poder18}. The next generation of multi-PW high-power lasers \cite{ELI,ELINP,corels19,omegaEP,shen18} (for a review, see \cite{danson19}), will allow a thorough exploration of this new regime.

    The two key strong-field QED processes to be investigated here are: the emission of a photon by an electron (or positron), known as \emph{nonlinear Compton scattering} (NLC) \cite{nikishov64,kibble64,narozhny65} and the decay of a high-energy photon into an electron-positron pair, known as the \emph{nonlinear Breit-Wheeler} (NBW) process \cite{reiss62,narozhny65}. For upcoming strong-field experiments it will be important not only to know the kinematic dependence and the particle spectra, but also the spin and polarisation dependency of these processes. First, because polarised high-energy electrons and photons find numerous applications such as nuclear spectroscopy \cite{horikawa14} and in being ideal probes for strong-field loop processes such as photon-photon scattering \cite{ilderton16,king16,nakamiya17,meuren17b}. Second, to correctly model the incoherent part of higher-order effects such as the trident process \cite{ritus72,narozhny77,hu10,ilderton11,king13b,torgrimsson18,mackenroth18,king18,torgrimsson20,torgrimsson20b} (creation of an electron-positron pair from a photon emitted by nonlinear Compton scattering) and double nonlinear Compton scattering \cite{morozov75,seipt12,mackenroth13,king15b,torgrimsson19}, the polarisation of the intermediate particle must be taken into account. This naturally poses the question of how important polarisation effects are in the modelling of strong-field electromagnetic cascades \cite{nerush11,mironov14,bashmakov14,gelfer15,narozhny15,tamburini17,grismayer17,zhang20,seipt20}, which are, in general higher than second order in multiplicity. Third, several processes in the extension of strong-field QED to beyond-the-standard-model physics, are sensitive to particular polarisation channels, such as axionic nonlinear Compton scattering, which, due to the emission of a pseudoscalar, only proceeds with a spin-flip \cite{borisov96,king2018electron,dillon2018alp}, and the decay of an axion into an electron-positron pair, which has a preference for the spin of the particles produced \cite{king2019axion}. 

    If the electromagnetic background is sufficiently weak, then spin and polarisation effects can be studied in perturbative QED. For \emph{linear} Compton scattering this has been done in many works, with Klein and Nishina already studying the effect of photon polarisation \cite{klein29}, and others looking at the role of spin and polarisation \cite{fano49,lipps54,bhatt83,ahrens17}. Similarly, the photon polarisation dependence of pair production in the collision of two photons was already considered in the seminal work of \citet{breit34}. However, if the intensity of the background is strong enough that on average more than one photon interacts with an electron, one must consider nonlinear QED processes, typically studied in a \emph{plane-wave background}.

    To investigate the relevance of the electron spin in the NLC and NBW processes several authors compared calculations for spin-1/2 Dirac particles with corresponding spin-0 Klein-Gordon particles \cite{ehlotzky09,boca12,villalba13,jansen16}. The effect of the electron being a spin-1/2 particle on the radiated light spectrum was also studied experimentally for the case of strong crystal fields (channeling radiation) \cite{kirsebom01}. The difference brought by the electron having a spin is that the spin---and its associated magnetic moment---can ``flip'' during the photon emission which is a quantum effect. In a quantum treatment of radiation emission the photon spectrum has contribution from both the electric charge and from the magnetic moment, and both of these contributions are present even if the incident particles are unpolarised and the final state polarisation remains unobserved. The contribution of spin-flips to NLC has been studied in a plane-wave pulse in comparison to classical radiation calculations \cite{krajewska13,krajewska14}. Note the spin can also flip in a laser background due to a non-radiative process (e.g. involving the (dressed) mass operator), which has been studied in \cite{baier76,meuren11}. The combination of radiative and non-radiative spin-flipping has been studied semiclassically in a constant magnetic field \cite{baier72b}.

    In NLC, the emitted photon polarisation has been studied for an electron in a constant crossed field \cite{ritus72,narozhny77,ritus85,king13a,king15b}, in a monochromatic plane-wave background \cite{tsai93} and recently in a plane-wave pulse \cite{tang20a,king20a}. The dependence of NLC on the incident electron spin has been calculated in \cite{ritus70,ritus72b} from the one-loop mass operator via the optical theorem (therefore yielding no information about final state polarisation properties). The spin-polarisation dependence of NLC in a monochromatic plane wave has been studied in \cite{bolshedvorsky00} without considering the photon polarisation, in \cite{ivanov04} including the photon polarisation, and in \cite{wistisen19} in a pulse. In \cite{seipt18b} the electron spin-polarisation (averaged over photon polarisation) for NLC in a short pulse has been investigated using the density matrix formalism where also the LCFA was calculated. For the case of a constant, homogeneous magnetic field, in which an electron produces (quantum) synchrotron radiation, there have also been several studies for the spin-polarised but photon-unpolarised case \cite{tsai73,jackson76}, and all particles polarised \cite{sokolov,ternov95}. In some works special emphasis was placed on radiation by the anomalous spin magnetic moment \cite{jackson76,bordovitsyn95}. For a review on spin-polarised particle beams in synchrotrons see e.g. \cite{mane05}.

    For NBW pair production, the effect of photon polarisation (but unobserved spin state of the pair) has been calculated in a monochromatic plane wave \cite{toll52,nikishov64,nikishov67,ritus72} and in a constant crossed field \cite{nikishov67,ritus70,king13a}. Similar calculations have been performed also for constant magnetic fields \cite{tsai74} and
    arbitrary constant electromagnetic fields \cite{katkov12}. The spin of electrons and positrons produced in NBW has also been studied for a monochromatic background \cite{tsai93}, and the completely polarised NBW cross sections in a strong linearly and circularly polarised monochromatic plane-wave have been calculated in \cite{ivanov05}. Numerical results for a pulsed plane wave were obtained in \cite{wistisen20}. Spin-resolved pair production in a strong field has been calculated also for various different field configurations (and production processes) \cite{dipiazza10,muller11,kohlfurst19,huang19}.

    In the rest frame of an ultra-relativistic charge, an arbitrary strong electromagnetic field ``looks'' like a crossed field (as shown by e.g. the Weizs{\"a}cker-Williams approximation \cite{weizsaecker34,williams34}). If the field is sufficiently intense, the length scale on which both NLC and NBW are ``formed'', is much shorter than the length scale of the shortest inhomogeneity in the laser pulse, namely its wavelength. Hence, the probabilities can be calculated using a ``locally constant'' field approximation (LCFA). The significance of strong-field quantum effects can be quantified using the \emph{quantum nonlinearity parameter}, $\chi$, which is defined
    for electrons and photons as $\chi_{e,\gamma} = |F_{\mu\nu} p_{(e,\gamma)}^\nu|/ (m E_{cr})$,
    where $F$ is the background field strength tensor, $p_{(e,\gamma)}$ is the probe particle momentum, and $E_\mathrm{cr} = m^2/|e|$ is the Sauter-Schwinger critical field of QED, with electron mass $m$ and charge $e<0$.
    By colliding a high-energy electron beam with an intense laser pulse it is possible to reach the regime where $\chi_e \sim  1$ \cite{luxe,e320}. 
    In the quantum regime the LCFA is valid if $\xi\gg1$ and $\xi^3/\chi\gg1$, where $\xi=(m/\kappa^0)(E/E_{cr})$, $E$ is the field strength, and $\kappa^0$ is the frequency of the background. The quantity $\xi$ has the meaning of an inverse Keldysh-type parameter. For practical purposes, and with $\chi\sim 1$, the LCFA can be considered a reasonable approximation for $\xi \gtrsim 10$ \cite{blackburn18,seipt18b}, despite its known limitations \cite{king19a,dipiazza18a}. Monte-Carlo sampling of the LCFA rates \cite{ritus85,harvey15,dipiazza18a,dipiazza19,king20lcfa} is the central method by which strong-field QED effects are included in high-intensity laser-plasma simulations \cite{elkina11b,ridgers14,gonoskov15,grismayer17}. Some polarised LCFA rates have been already implemented in (Monte Carlo) simulation codes to investigate the radiative self-polarisation of fermions in different field configurations \cite{delsorbo17,delsorbo18,seipt19,chen19,li19} and to model photon polarisation effects \cite{king13a,li20}, as well as polarised QED cascade formation \cite{seipt20}.

    A reasonable amount of work has already been performed in investigating the role of polarisation and spin in different processes and different electromagnetic backgrounds. Yet, a systematic study of all spin and polarisation effects of the NLC and NBW processes and a consistent derivation of the LCFA is still lacking. This is achieved in this paper. The results for the completely polarised LCFA rates presented in this paper are suitable for a direct implementation in such numerical frameworks.
    In the current paper we present compact analytical expressions for the fully polarised NLC and NBW processes. We calculate these processes in a plane-wave pulse, from which the LCFA is derived. 
    We find strong correlations in the spectra between the polarization states of photons and leptons, especially when $\chi$ is large.
    Asymptotic formulas for the fully polarised rates are given for small and large values of the seed particle's quantum parameter, $\chi$, and compared quantitatively to the full LCFA.
    These asymptotic approximations are compared quantitatively to the full LCFA which shows some unexpectedly slow convergence for some particular polarization channels.
    All polarisation channels in each of the processes are visualised for various quantum parameter, and the relative ordering of each channel is explained phenomenologically. 
    
    The paper is organised as follows. In Sec. II we introduce the polarisation and spin bases and give an overview of the kinematics, crossing symmetry and general structure of the probabilities. Secs. III and IV present the results for NLC and NBW respectively. Both sections include a presentation of the results for a plane wave and for the LCFA, for which the spin polarised asymptotic scaling for large and small quantum nonlinearity parameter is given. Noteworthy aspects of the results are discussed at the end of each section. In Sec. III, we also include an overview of the derivation. In Sec. V the paper is concluded. Appendices A and B contain a detailed derivation of the results for NLC and NBW respectively. Throughout the paper we employ natural Heaviside-Lorentz units with $\hbar=c=\varepsilon_0=1$.

    \section{Polarisation Basis}
    
    We begin by introducing the polarisation states of vector and spinor particles that will be appear throughout the calculation and in our final results for the fully polarised nonlinear Compton (NLC) and nonlinear Breit-Wheeler (NBW) rates. We will concentrate on the case of a linearly polarised plane-wave laser pulse of arbitrary temporal shape. We introduce the laser polarisation $\varepsilon^\mu$ and four-wavevector $\kappa^\mu$, satisfying $\varepsilon.\varepsilon=-1$, $\kappa.\kappa=0$ and $\varepsilon.\kappa=0$. 
    The normalised vector potential of the background, $a=eA$ with $e<0$, depends only on the phase variable $\phi = \kappa.x$, and can be given by $a^\mu(\phi) = m\xi \varepsilon^\mu h(\phi)$, with the classical nonlinearity parameter $\xi$ and an arbitrary shape function $h(\phi)$.
    In addition, it is useful to define the constant background field tensor $f^{\mu\nu} = \kappa^\mu \varepsilon^\nu - \kappa^\nu \varepsilon^\mu$.  Let us also define the magnetic field polarisation, $\beta$, satisfying $\beta.\beta=-1$, $\beta.\varepsilon=\beta.\kappa=0$.
    The spatial components of the four-vectors ($\varepsilon,\beta,\kappa$) need to form a right-handed triad. 
	For instance, in the lab frame we can choose $\kappa = \omega(1,0,0,1)$, $\varepsilon=(0,1,0,0)$, $\beta = (0,0,1,0)$, where $\omega$ is the laser frequency.
	This ensures that their spatial parts fulfill $\vec \kappa /\omega = \vec \epsilon\times \vec \beta$, i.e. $\vec \kappa$ agrees with the direction of the background field Poynting vector.

    \subsection{Photon polarisation Basis}
    
    With the triad of basis vectors $(\varepsilon,\beta,\kappa)$, we can  define a (linear) polarisation basis for a photon with four-momentum $k$ as
    \bea
     \Lambda_{1} = \eps - \frac{k .  \eps}{k. \kappa} \kappa \,; \qquad
    \Lambda_{2} = \beta - \frac{k . \beta}{k . \kappa} \kappa \,. 
    \eea
    By construction the polarisation basis vectors fulfil $k.\Lambda_j=0$ and $\Lambda_i.\Lambda_j = -\delta_{ij}$.
    An arbitrarily polarised photon (in a pure state) with polarisation four-vector $\epsilon_k$ can therefore be written as the superposition
    \begin{align}
    \epsilon_k = c_{1}\Lambda_{1} + c_{2}\Lambda_{2} \,.
    \end{align}
    We will characterise the photon polarisation state using the Stokes parameter $\tau_k = |c_1|^2 - |c_2|^2$, where $\tau_{k}$ is, in general, a real number and $\tau_{k}\in [-1,1]$. We note that, if $\tau_{k}$ is chosen to be an integer and $\tau_{k}\in\{-1,1\}$, then the photon is produced in an eigenstate of the polarisation operator \cite{baier75a}, and therefore the polarisation will not precess as the photon propagates through the background (reviews of photon-photon scattering can be found in \cite{marklund06,dipiazza12,king15a}). In this paper, we will consider the case that $\tau_{k}\in\{-1,0,1\}$, where $\tau_k=0$ indicates an unpolarised photon (in a mixed state), i.e.~a polarisation average. (For unobserved final state polarisation one has to multiply the result by 2.) A photon in the $\Lambda_1$ polarisation state is polarised parallel to the laser polarisation direction in a frame in which $k$ and $\kappa$ are collinear. A photon in the $\Lambda_2$ polarisation state is polarised perpendicular to the laser.
    Hence, we may refer to these photons as $\parallel$- and $\perp$-polarised photons, respectively.

    \subsection{Fermions}
    
    The spin basis can be chosen in a similar way to the photon polarisation basis. It is useful to define a basis that does not precess in the background field. Also, the basis cannot depend on spacetime co-ordinates, otherwise we would be modifying the spacetime dependency of the Volkov solution, which would not fulfill the Dirac equation anymore. 
    For \emph{linearly-polarised backgrounds} in the $\eps$ direction our basis for the spin four-vector of an electron with momentum $p$ becomes: 
    \bea
    \zeta_{p} &=& \beta - \frac{p . \beta}{p .  \kappa} \kappa \,.
    \eea
    Then we see that $\zeta_{p} . \zeta_{p} = -1$ and $\zeta_{p} .  p = 0$, but also very usefully: $\zeta_{p} . \kappa =0$
    and $\zeta_p.\eps=0$.
    The choice of this basis vector implies that we are looking specifically at \emph{light-front transverse} polarisation, with the spin-vectors
    oriented along the magnetic field in the rest frame of the particle.
    An important aspect of this choice of the spin-quantisation axis is that then $F . \zeta_{p} = 0$, where $F$ is the background field strength tensor. This fact immediately ensures that the spin-vector of the particles does not precess under the Bargman-Michel-Telegdi (BMT) equation \cite{bargmann59},
    \begin{align}
        \frac{\ud S^\mu}{\ud \tau } &= \frac{eg_e}{2m} F^{\mu\nu} S_\nu  - \frac{e(g_e-2)}{2m} u^\mu \, (u.F.S) \,,
    \end{align}
    where $S^\mu$ is a general spin-polarisation vector, $g_e$ the electron gyromagnetic ratio, and $u^\mu$ its four-velocity. Although $\zeta_{p}$ is defined using the asymptotic momentum, $p$, we see that we can replace, without loss of generality, $p$ with the ``instantaneous'' classical kinetic momentum $\pi_{p} (= m u)$ of the electron in a plane-wave background,
	\begin{align}
		\pi_p(\phi) = p - a + \kappa \frac{p.a}{\kappa.p} - \kappa \frac{a.a}{2\kappa.p} \,, \label{eqn:pi1}
	\end{align}
    and hence $\zeta_{\pi} \equiv \zeta_{p}$.

    The choice of the basis above $S^\mu =\zeta^\mu$ therefore ensures that $\ud S^\mu /\ud \tau =0$. Thus, the asymptotic polarisation state of the particles agrees with the local values inside the strong background field.
    This is a special choice of spin basis. In general one could expand the spin vectors in a dreibein: $S^\mu = S_\zeta \zeta_p^\mu + S_\eta \eta_p^\mu + S_\varkappa \varkappa_p^\mu$, where $\eta_p$ and $\varkappa_p$ are two additional space-like unit four-vectors perpendicular to $p$, and defined as
    $\eta_p^\mu = \varepsilon^\mu - \kappa^\mu (p.\varepsilon)/(p.\kappa) $ and $\varkappa_p^\mu = m \kappa^\mu /(\kappa.p) - p^\mu/m$, (noting $F.\eta \neq 0$ and $F.\varkappa\neq0$). Thus, the BMT equation would imply that a general spin vector precesses. It can be shown that the vectors ($\zeta_p,\eta_p,\varkappa_p$) are pointing in the direction of the background magnetic field, electric field, and wave-vector in the rest frame of the particle \cite{seipt18b}. 

     The Dirac bi-spinors are defined using the spin basis $\zeta_p$, which is manifest in the density matrices \cite{itzykson}:
    \begin{align}
	u_{p \sigma_p} \bar u_{p \sigma_p} &= \frac{1}{2} (\slashed p  + m ) (1 + \sigma_p \gamma^5 \slashed \zeta_p ) \,,\\
	v_{p\sigma_p} \bar v_{p\sigma_p} &=  \frac{1}{2}(\slashed p - m) (1 + \sigma_p \gamma^5 \slashed \zeta_p ) \,,
	\end{align}
    where we explicitly introduce the spin index $\sigma_p = \pm1$ to distinguish states where the spin vector is parallel (spin-$\uparrow$, $\sigma_p=+1$) or anti-parallel (spin-$\downarrow$, $\sigma_p=-1$) to $\zeta_p$.

    \subsection{General considerations for the polarisation-resolved probabilities}

    \begin{figure}[h!!]
        \centering
        \includegraphics[width=\columnwidth]{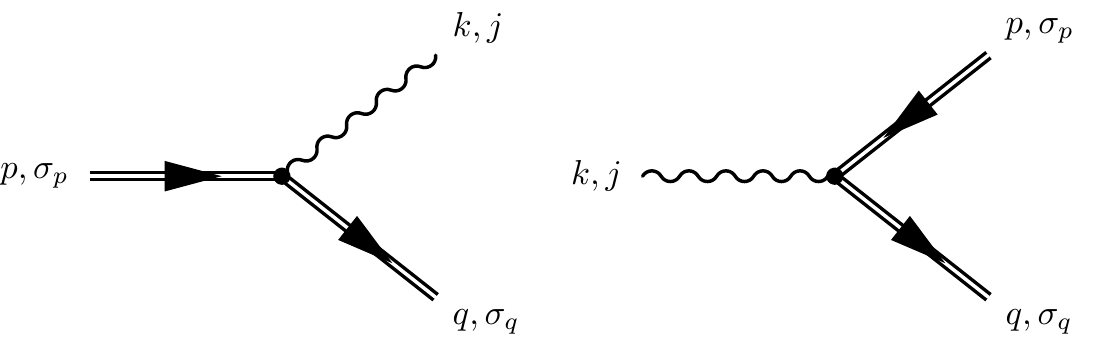}
        \caption{Feynman diagrams. Left: nonlinear Compton scattering (NLC). Right: nonlinear Breit-Wheeler (NBW) pair production.}
        \label{fig:feynman}
    \end{figure}

    Nonlinear Compton scattering (NLC) and nonlinear Breit-Wheeler (NBW) pair production are both $1\to2$ first-order strong-field QED processes with one interaction vertex (see \figref{fig:feynman}). Their corresponding S-matrix elements are related by crossing invariance. The strong-field QED vertex is an interaction of two ``dressed'' fermion lines (including the exchange of a number of background laser photons) and one photon line. We preface the detailed calculation of these two processes with some general remarks on the (light-front) kinematics of the processes and the structure of the expressions for the polarisation-resolved probabilities.

    To have a unified notation for both processes under study, let us denote the incoming momentum as $p_\mathrm{in}$, the outgoing momentum of the particle under study
    as $p_\mathrm{out}$, and the momentum of the outgoing particle we integrate over (i.e. its momentum is not observed, but its polarisation state is) as the ancillary momentum $q$. We orient the coordinate system in such a way that the laser propagates along the \emph{positive} $z$-axis, i.e.~$\kappa^+ =2\omega$ (where light-front momentum components are defined $p^\pm=p^{0}\pm p^{3}$) is the only non-vanishing light-front component of $\kappa^\mu$. Then, for both processes, NLC and NBW, the light-front momentum conservation can be expressed as
    \begin{align} \label{eq:emc-lf}
    p_\mathrm{in}^- = p_\mathrm{out}^- + q^- \,, \qquad
    \vec p_\mathrm{in}^\perp = \vec p_\mathrm{out}^\perp + \vec q^\perp \,,
    \end{align}
    with $\vec p^\perp = (p^1,p^2)$, and the exchange of ``$+$'' momentum between the particles and the background field does not yield a conservation law.
    In a plane-wave background, one can write the S-matrix element 
    using the four-dimensional light-front delta function 
    $\delta^{(4)}(P+\ell \kappa ) = 2 \delta(P^+ + \ell \kappa^+ ) \delta(P^-)\delta^{(2)}(\vec P^\perp) = 2 \delta(P^+ + \ell\kappa^+) \delta_\mathrm{l.f.}( P )$ as:
	\begin{align}
    	S&= -ie (2\pi)^4 \int \frac{\ud \ell}{2\pi} 
    	\delta^{(4)}(P + \ell \kappa   ) \: \mathscr M \,, 
	\end{align}
    where $P = p_\mathrm{in} - p_\mathrm{out} - q$, and the integral over $\ell$ takes into account exactly the non-conservation of $+$-momentum.
    The amplitude $\mathscr M$ is specific to each process, and contains all the spin- and polarisation dependence. The phase-space integrated probability for the process under consideration is then given by
    \begin{align}
        \mathbb P 
                & = 
                \frac{1}{2p_\mathrm{in}^-} \int \! \widetilde{ \ud^3 q} \,  
                \widetilde{\ud^3 p}_{\mathrm{out} } \: |S|^2  =: \int \! \ud \Gamma \: |S|^2
    \end{align}
    with the Lorentz-invariant on-shell phase space elements understood in light-front coordinates, i.e.~$\widetilde{\ud^3 q } = \frac{\ud q^- \ud^2 \vec q^\perp }{(2\pi)^2 2 q^-}$.

    The conservation of three light-front momentum components in Eq.~\eqref{eq:emc-lf} allows one to completely integrate out the ancillary momentum $q$. The final particle phase space of $\vec{p}_\mathrm{out}$ is conveniently parametrised by the normalised light-front momentum transfer $s$ and transverse momentum $\vec r^\perp$:
    \bea
     s :=  \frac{p_\mathrm{out}^-}{p_\mathrm{in}^-} =  \frac{\kappa.p_\mathrm{out}}{ \kappa. p_\mathrm{in} }  \,, \qquad
     \vec r^\perp := \frac{\vec p_\mathrm{out}^\perp}{ms} \,.
    \eea

    We thus can write the final particle phase space as
    \begin{align}
	    \ud \Gamma =  
	        \ud q^- \ud^2 \vec q^\perp \left(\frac{m}{p^-}\right)^2 \frac{s}{1-s} \frac{\ud s\, \ud^2 \vec r_\perp}{8 (2\pi)^6} \,.
    \end{align}
    Moreover, for the squared S matrix we find
    \begin{align}
        |S|^2 &= (2\pi)^3 e^2 \left( \frac{2}{\kappa^+} \right)^2 \delta_\mathrm{l.f}\left(P\right)\: |\mathscr M|^2,
    \end{align}
    where we used the normalisation $\delta_\mathrm{l.f.}(0) = \frac{1}{(2\pi)^3}$.
    Integrating out the ancillary momentum $q$ consumes the delta function and allows the total probability to be expressed as 
    \begin{align} \label{eq:probability-general}
        \mathbb P  & = 
        \frac{ \alpha  }{16\pi^2  m^2 b^2 } 
        \int_0^1 \frac{\ud s \,s }{1-s } \: \int \! \ud^2 \vec r_\perp   \: |\mathscr M |^2 \,,
    \end{align}
    with fine structure constant $\alpha = e^2/4\pi$, and quantum energy parameter $b=p_\mathrm{in}.\kappa /m^2$. The squared amplitude is given by a double integral over the laser phase, which takes the form
    \begin{align} \label{eq:sqared-amplitude-general}
        |\mathscr M|^2 &= \int \ud \phi \,\ud \phi' \: e^{i \Phi } \: 
    \mathsf T_j \,.
    \end{align}
    Here, the integrand is a product of a nonlinearly-oscillating factor, the trace of Dirac matrices 
    $\mathsf T_j = \Lambda_j^\mu  \mathsf T_{\mu\nu} \Lambda_j^\nu$ containing the fermion spin structure in $\mathsf T_{\mu\nu}$, and photon polarisation vectors $\Lambda_j^\mu$.
    The specific form of these expressions depends on the considered process. In the following sections, they are evaluated in a linearly polarised plane wave laser background, first for nonlinear Compton scattering and then for nonlinear Breit-Wheeler pair production. From the general plane-wave results, we then rigorously derive the locally constant field approximation.

    \section{Nonlinear Compton Scattering}

    This section is devoted to the investigation of fully polarised nonlinear Compton scattering, i.e.~the emission of a polarised photon by a spin-polarised electron, where also the spin-polarisation after the photon emission is observed. We restrict the discussion to the case of all particles being in polarisation eigenstates as discussed above:
    Initial (final) electrons can be spin-polarised $\sigma_p=\pm1$ ($\sigma_q=\pm1$) along the axis $\zeta_p$ ($\zeta_q$);
    photons are emitted in polarisation eigenstate $\Lambda_j$, $j=1,2$.

    \subsection{S-Matrix}

    We begin by recalling the basic properties of 
	Volkov states, which are solutions of the Dirac equation in a plane-wave background,
	\begin{align} \label{eq:dirac}
	  ( i \slashed \partial - e \slashed A - m ) \Psi_{p \sigma_p}(x) = 0 \,,
	\end{align}
	and, with the normalised vector potential $a=eA$,
	given by
	\begin{align}
	\Psi_{p \sigma_p}(x) &= E_p(x)  u_{p\sigma_p} \,, \\
	E_p(x) &= \label{eq:ritusmat}
	\left( 1 + \frac{  \slashed \kappa \slashed a }{2p.\kappa } \right)
	\exp
	    \left\{ 
	        -ip.x - \int \! \d \phi \: \frac{2 a.p-a.a}{2 \kappa . p} 
	        \right\} \,,
	\end{align}
	where $E_p$ are the ``Ritus matrices'', $u_{p\sigma_p}$ are the Dirac bi-spinors, and where
	$\sigma_p = \pm1$ means the electrons are asymptotically aligned/anti-aligned with the space-like spin-quantisation axis $\zeta_p$. 
	Because $\zeta_p=\zeta_\pi$ they remain polarised in that state during the interaction with the laser prior to emitting a photon --- and after.

	Let us recall that the normalised vector potential $a$ of the background plane wave depends only on the phase variable $\phi = \kappa.x$, and is represented by  $a^\mu(\phi) = m\xi \varepsilon^\mu h(\phi)$, where $\xi$ is the classical nonlinearity parameter \cite{ritus85}, $\varepsilon^\mu$ is the polarisation vector obeying $\varepsilon.\varepsilon=-1$, and $h(\phi)$ is an arbitrary shape function. Examples of shape functions include $h(\phi)=\cos\phi$, for a linearly polarised infinite plane wave; and $h(\phi)=\phi$ for a constant crossed field. We now write for the (normalised) field strength tensor $F^{\mu\nu} = m\xi f^{\mu\nu} \dot h(\phi)$, where $f^{\mu\nu} = \kappa^\mu \varepsilon^\nu - \kappa^\nu \varepsilon^\mu$ is a constant tensor and	$\dot h (\phi) = \ud h/\ud \phi$.

	The S-matrix element for this strong-field QED process, see \figref{fig:feynman} left, reads
	\begin{multline}
	S_\mathrm{NLC}(\sigma_p,\sigma_q,j) = -ie \int \! \ud^4 x \: \bar \Psi_{q\sigma_q}(x) \,  \slashed \Lambda_j e^{ik.x}\,  \Psi_{p\sigma_p}(x)   \\
	 = -ie (2\pi)^4 \int \frac{\ud \ell}{2\pi} \delta^{(4)}(p+\ell \kappa - q - k ) \: \mathscr M_\mathrm{NLC}
	\end{multline}
	with the amplitude 
	\begin{multline}
	\mathscr M_\mathrm{NLC}(\sigma_p,\sigma_q,j) = \\ \Lambda_{j,\mu} \int \! \ud \phi \: e^{  i \int d\phi \frac{k.\pi_p(\phi) }{\kappa. q}} \: 
	\bar u_{q\sigma_q}  \mathscr J^\mu_\mathrm{NLC} (\phi) u_{p\sigma_p} \,,
	\end{multline}

	and the Dirac current which is independent of the polarisation properties of all particles
	\begin{align} \label{eq:current.NLC}
    \mathscr J^\mu_\mathrm{NLC}(\phi) = \gamma^\mu 
    +  \frac{\slashed a \slashed \kappa \gamma^\mu }{2(\kappa.q)} +  \frac{ \gamma^\mu \slashed \kappa \slashed a}{2(\kappa. p)}
    + \frac{\slashed a \slashed \kappa \gamma^\mu \slashed \kappa \slashed a}{4(\kappa.p)(\kappa.q)} \,.
    \end{align}

    \subsection{NLC Probability}

    With the results from Eq.~\eqref{eq:probability-general} we can write the probability as:
    \begin{multline} \label{eq:probability}
    \mathbb P_{\mathrm{NLC},j}(\sigma_p,\sigma_q)   = \\ 
        \frac{ \alpha  }{16\pi^2  m^2 b_p^2 } \int_0^1 \frac{\ud s \,s }{1-s } \: \int \! \ud^2 \vec r_\perp   \:
        |\mathscr M_\mathrm{NLC}(\sigma_p,\sigma_q,j)|^2 \,,
    \end{multline}
    where the squared amplitude is given by a double phase integral
    over a dynamic phase factor, which is independent of the particle polarisation, multiplied by $\mathsf T_j$, which is the Dirac trace $\mathsf T^{\mu\nu}$, contracted with the outgoing photon polarisation vectors, $\mathsf T_j = \Lambda_{j}^{\mu} \mathsf T_{\mu\nu}(\sigma_p , \sigma_q) \Lambda_j^\nu$. Explicitly,
    \begin{align}
    |\mathscr M_\mathrm{NLC}(\sigma_p,\sigma_q,j)|^2
			= \int \ud \theta\, \ud \varphi \: e^{i \theta \: \frac{k .\langle \pi_p \rangle }{\kappa.q}} \: 
    \mathsf T_j \,,
    \end{align}
    with $\theta = \phi-\phi'$, $\varphi = (\phi+\phi')/2$, and the floating average defined by
    \begin{align}
    \langle \pi_p \rangle = \langle \pi_p \rangle(\varphi,\theta)
    = \frac{1}{\theta} \int_{\varphi-\theta/2}^{\varphi+\theta/2} \! \d \phi'' \: \pi_p(\phi'') \,.
    \end{align}
    The dynamic phase for Compton scattering is given by
    \begin{align}
    \theta \frac{k . \langle\pi_p\rangle }{\kappa .q }
    &= \frac{ s\theta }{2b_p (1-s) } [ \mu  + ( \vec r^\perp - \langle \vec \pi_p^\perp \rangle /m)^2]
    \end{align}
    with normalised Kibble's mass
    \begin{align} \label{eq:Kibblemass}
     \mu = 1 + \xi^2 \langle h^2 \rangle - \xi^2  \langle h \rangle^2,
    \end{align}
    energy parameter $b_p = \kappa.p/m^2$, and $s=\kappa.k/\kappa.p$. The spin trace
    \begin{multline}
    \mathsf T^{\mu\nu} 
                = 
                \frac{1}{4} \mathrm{tr} \, \Big[
                (\slashed q + m ) (1+\sigma_q \gamma^5 \slashed \zeta_q) \mathscr J^\mu(\phi) \\
                \times
                 (\slashed p + m ) (1+\sigma_p \gamma^5 \slashed \zeta_p) \bar{  \mathscr J}^\nu(\phi ' )
            	\Big] \,.
    \end{multline}
    can be decomposed into four parts: unpolarised ($\UP$),
    initially polarised ($\IP$, depends only on the initial electron polarisation), finally polarised ($\FP$, depends only on the final electron polarisation), and polarisation correlation ($\PC$, depends on both the initial and final electron polarisation).
    These terms are defined as follows:
    \begin{align}
    \mathsf T^{\mu\nu} (\sigma_p,\sigma_q) & = 
     \UP^{\mu\nu} + \sigma_p \IP^{\mu\nu} + \sigma_q \FP^{\mu\nu} + \sigma_p \sigma_q \PC^{\mu\nu} \,,
    \end{align}
    with the four contributions
    \begin{align} \label{eq:NLCtraceUP}
     \UP^{\mu\nu} & \equiv
    	\frac{1}{4} \TR{ (\slashed q + m ) \: \mathscr J^\mu(\phi) \: (\slashed p + m) \: \bar{ \mathscr J }^\nu(\phi') 
    	    \vphantom{\slashed \zeta_q} }  \,, \\
     \FP^{\mu\nu} & \equiv
    	\frac{1}{4} \TR{  (\slashed q + m ) \: \gamma^5\slashed \zeta_q \: \mathscr J^\mu(\phi)  \: (\slashed p + m ) \: \bar{ \mathscr J }^\nu(\phi') } \,,  \\
     \IP^{\mu\nu}  & \equiv
    	\frac{1}{4} \TR{  (\slashed q + m ) \: \mathscr J^\mu(\phi) \:  (\slashed p + m )\: \gamma^5\slashed \zeta_p \: \bar{ \mathscr J }^\nu(\phi') } \,, \\
     \PC^{\mu\nu} & \equiv
    	\frac{1}{4} \TR{   (\slashed q + m ) \: \gamma^5\slashed \zeta_q \: \mathscr J^\mu(\phi) \: 
    	 							(\slashed p + m ) \: \gamma^5\slashed \zeta_p \: \bar{ \mathscr J }^\nu(\phi') } \,,
    	 							\label{eq:NLCtracePC}
    \end{align}
    {where the NLC current from Eq.~\eqref{eq:current.NLC} and its Dirac-adjoint $\bar {\mathscr J} = \gamma^0 \mathscr J^\dagger \gamma^0$ have to be inserted. (\textsc{FeynCalc} \cite{feyncalc,feyncalc93} was used to calculate the traces.) Then, the expression for the differential probability, can be written as
    \begin{multline} \label{eq:Pj_NLC}
    \frac{\ud \mathbb P_{\mathrm{NLC},j}}{\ud s} (  \sigma_p , \sigma_{q} ) = \frac{\alpha}{16\pi^2 m^2 b_p^2 } \frac{s}{1-s} \: \int \! \ud \varphi
		 \int \! \ud \theta \int \! \ud^2 \vec r_\perp \\
	     \times \: e^{i \theta \: \frac{k .\langle \pi_p \rangle }{\kappa.q}}
		 \left[  \UP_j + \sigma_p \IP_j + \sigma_{q} \FP_j +  \sigma_p \sigma_{q}  \PC_j \right] \,,
    \end{multline}
    for photons emitted in a polarisation state $j=1,2$. Introducing the Stokes parameter $\tau_k$ of the emitted photon then yields
    \begin{align}
    \frac{\ud \mathbb P}{\ud s}( \sigma_p ,  \sigma_{q} , \tau_k ) 
		& = 
			\frac{ 1 + \tau_k }{2}  \frac{\ud \mathbb P_1}{\ud s}
		  + \frac{ 1-\tau_k }{2}  \frac{\ud \mathbb P_2}{\ud s} \,.
    \end{align}

    After evaluating the total of eight different traces, and analytically performing the integration in $\vec r _{\perp}$, which is Gaussian (the technical details of these steps are giving in Appendix~\ref{app:nlc}), and regularizing the resulting expressions (e.g. with an ``$i\varepsilon$'' prescription \cite{dinu13}), one arrives at the expression for the NLC spectrum in a \emph{plane-wave pulse}:
    \begin{widetext}
    \begin{align} \label{eqn:pulse1}
    \frac{\ud \mathbb P_\mathrm{NLC}}{\ud s} & = -\frac{\alpha}{8\pi b_p} \int \! \ud \varphi 
    \int \! \frac{\ud \theta}{-i\theta} \: e^{i x_0 \theta \mu} \: \mathcal I_\mathrm{NLC} \,,  \\
    \mathcal I_\mathrm{NLC}     &= 1 + \sigma_p\sigma_q +(1-g)\tau_k\sigma_p\sigma_q  
    -\xi^2 \Delta h^2 \tau_k (1+g\sigma_p\sigma_q)
    + \frac{\xi^2 \langle\dot h\rangle^2 \theta^2 }{2} (g+\sigma_p\sigma_q )
    \nonumber \\
    & \qquad - \frac{i\theta \xi \langle \dot h\rangle }{2}
    \left[
        s \sigma_p + \frac{s}{1-s} \sigma_q
        +\tau_k \left( s\sigma_q + \frac{s}{1-s} \sigma_p\right)
    \right] \,,
    \end{align}
	\end{widetext}
    where $\Delta h^2 = (h(\phi)-\langle h \rangle)(h(\phi')-\langle h \rangle)$, 
    $x_0=s/[2b_p(1-s)]$ and $g=1+s^2/[2(1-s)]$.    A numerical evaluation of this expression calls for an additional regularization of that part of $\mathcal I_\mathrm{NLC}$ not containing the laser pulse, i.e.~being $\propto \xi^0$. Several methods for this regularization have been discussed in the literature \cite{boca09,torgrimsson18,heinzl20}.

    The appearance of a pre-exponential term proportional to $1/\theta^2$, see e.g. Eqs.~\eqref{eq:UP1-theta21} and \eqref{eq:UP1-theta22}, is known from polarised calculations in a plane wave \cite{king20a}. 
    In the expression above it has already been treated using integration by parts, giving terms 
    \begin{align}
    \frac{\ud (\theta\mu)}{\ud\theta} 
    = 1 + \xi^2 \Delta h^2 + \frac{\theta^2 \xi^2 \langle \dot h \rangle^2 }{2} \,.
    \end{align}

    To acquire the LCFA, and specifically a \emph{local rate}, one performs an expansion of the exponent in Eq.~\eqref{eqn:pulse1} to cubic order in $\theta$ and each term in the pre-exponent to leading order $\theta$. Then, the integrals over $\theta$ can be performed analytically. Let us define the probability rate 
    $ \mathbb R  = \ud \mathbb P / \ud \varphi $ as the probability for emission per unit laser phase. Combining \eqref{eq:RNLC1} and \eqref{eq:RNLC2}, the differential NLC rate for all particles polarised is then given by
    \begin{widetext}
    \begin{align}
    \frac{\ud \mathbb R_\mathrm{NLC}}{\ud s}( \sigma_p ,  \sigma_{q} , \tau_k ) 
		 &=  -  \frac{\alpha}{4b_p} 
		  \left[ 
		 \{ 1+\sigma_p\sigma_q+\tau_k\sigma_p\sigma_q (1-g) )\} \: \Ai_1(z)  \vphantom{\frac{1}{1}}\right. \nonumber \\
		 & \qquad \quad 
		 		+ \left\{ s\sigma_p + \frac{s}{1-s} \sigma_q +\tau_k \left( \frac{s}{1-s}\sigma_p + s\sigma_q \right) \right\} \frac{\Ai(z)}{\sqrt{z}} \: \shdp	
		 			    \nonumber\\
		&   \left.
				\qquad \quad + \left\{  g+\sigma_p\sigma_q +\tau_k \frac{1+g\sigma_p\sigma_q}{2} \right\} \frac{2\Ai'(z)}{z}
		  \right]. \,
		  \label{eq:RNLCfinal}
    \end{align}
    \end{widetext}
    The argument of the Airy function $\Ai(\cdot )$, its derivative $\Ai'(\cdot)$ and integral 
    $\Ai_1(z) := \int_z^\infty \ud x \, \Ai(x)$ is $z = (\frac{s}{\chi_e(\varphi) (1-s)})^{2/3}$ and depends on the local value $\chi_e(\varphi) = \chi_p |\dot h(\varphi)|$, where $\chi_p = \xi b_p$. 
    The term $\shdp$ in the second line of \eqref{eq:RNLCfinal} appears because of the oscillating nature of a plane wave pulse. It shows that this particular term switches its sign each half cycle of the wave together with the direction of the magnetic field. Hence,
    in an oscillating field with many cycles one can expect that this term averages to zero when integrating the rate over the pulse if the field has a certain symmetry such that, integrated over a cycle $\int \!\ud  \varphi \: \shdp \Ai(z)/\sqrt{z} \approx 0$.
    Since $z$ only depends on $|\dot h|$ this is the case if the field has some (generalised) parity property $\dot h ( \phi_0\pm\phi ) \approx - \dot h (\phi)$
    for some $\phi_0$. In order to efficiently radiatively polarise electrons this symmetry needs to be broken, for instance using an ultra-short sub-cycle pulse \cite{seipt18b}, or
    by a bi-chromatic (two-color) field \cite{seipt19}. By superimposing a 2nd harmonic with the correct phase, e.g.~$\dot h = \cos \phi + \cos 2\phi $, the (generalised parity) symmetry is broken and it is impossible to find a $\phi_0$ such that $-\dot h(\phi) \approx \dot h(\phi_0 \pm \phi)$. Similar arguments also hold for NBW pair production \cite{chen19}.

From this expression we can straightforwardly recover literature results for the partially polarised cases.
The case for unobserved photon polarisation is acquired by setting $\tau_k=0$ and multiplying the result by 2 (for the sum over the final polarisation states)
\begin{multline}
2\,\frac{\ud \mathbb R_\mathrm{NLC}}{\ud s}( \sigma_p ,  \sigma_{q} , \tau_k = 0 ) 
=  -  \frac{\alpha}{2b_p} 
\left[ 
	(1+\sigma_p\sigma_q) \Ai_1(z) \vphantom{\frac{1}{1}}
	\right. \\
	+ \left(  s\sigma_p+ \frac{s}{1-s} \sigma_q  \right) \frac{\Ai(z)}{\sqrt{z}} \: \shd \\
		\left. +  (g+\sigma_p\sigma_q ) \frac{2\Ai'(z)}{z}
\right] .
\end{multline}
This result agrees with the diagonal elements of the spin-density matrix in Ref.~\cite{seipt18b}. 

The rate for unpolarised final state particles, but polarised initial electrons, had been calculated, e.g.
by Ritus via the imaginary part of the one-loop electron mass operator
\cite{ritus70,ritus72b}.
We can obtain this from the general expression by setting $\sigma_q=\tau_k=0$ and multiplying by 4 to take into account the summation over final state particles 
    \begin{multline} 
    4 \, \frac{\ud \mathbb R_\mathrm{NLC}}{\ud s}( \sigma_p ,  \sigma_{q} = 0 , \tau_k = 0 ) 
    =  \\ 
    -  \frac{\alpha}{b_p} 
    \left[  \Ai_1(z)
	+  s \sigma_p \, \frac{\Ai(z)}{\sqrt{z}} \: \shd
	 +  g \, \frac{2\Ai'(z)}{z}
    \right] .
    \end{multline}

Finally, the case of unpolarised electrons, but polarised photons can be achieved by setting $\sigma_p=\sigma_q=0$ and multiplying by $2$ (which is equivalent to performing an average over incoming spins and a sum over outgoing ones) to achieve:
    \begin{multline} \label{eq:NLCspinunpol}
    2 \, \frac{\ud \mathbb R_\mathrm{NLC}}{\ud s}( \sigma_p = 0,  \sigma_{q} = 0 , \tau_k  ) 
    =  \\ 
    -  \frac{\alpha}{2b_p} 
    \left[ 
    	\Ai_1(z) +  (2g+\tau_{k}) \frac{\Ai'(z)}{z}
    \right] ,
    \end{multline}
    which agrees with literature results \cite{king13a}.

    Finally, the completely unpolarised nonlinear Compton rate
    is obtained by setting $\sigma_p=\sigma_q=\tau_k=0$ and multiplying by $4$ for the summation over the final electron spin and photon polarisation states, yielding \cite{kibble64,nikishov64}.
    \begin{multline} \label{eq:NLCunpol}
      4 \, \frac{\ud \mathbb R_\mathrm{NLC}}{\ud s}( \sigma_p = 0,  \sigma_{q} = 0 , \tau_k =0 ) 
    = \\
    -  \frac{\alpha}{b_p} 
    \left[ 
    	\Ai_1(z) +  2g \frac{\Ai'(z)}{z}
    \right] .
    \end{multline}

    We can also make a connection to the expressions calculated by Sokolov and Ternov in a constant and homogeneous magnetic field. Translating the Airy functions into modified Bessel functions of the second kind and setting $\dot h =1$ we get perfect agreement with the expressions from the literature \cite{sokolov}.

\subsection{Discussion of the Compton Rates}

    To discuss the relative and absolute weight of the eight different polarisation channels, we plot the different NLC emission rates for a constant value of $\chi_e=\chip$ in Fig.~\ref{fig:PEtotal}.
    We can make the following general remarks. 
    For the total yield of photons due to each polarisation channel, shown in \figref{fig:PEtotal}, we see the channels without a spin-flip are much larger than those with a spin-flip. The dominant contribution is the non-flip transition when the polarisation of the emitted photon is in the $\parallel$ state (which is approximately parallel to the background electric field for a near head-on collision of electron and laser pulse). The non-flip channels with the photon emitted in the $\perp$ polarisation state are next in the hierarchy of rates.
    All spin-flip rates are much lower than the non-flip rates. Especially for $\chip \ll 1$ they are suppressed by additional powers of $\chip$ (c.f.~the discussion of the asymptotic behaviour below). The most probable spin-flip channel is the emission of a perpendicularly polarised photon during an $\uparrow$ to $\downarrow$ transition.

    \begin{figure}[!bht]
    \centering
    \includegraphics[width=\columnwidth]{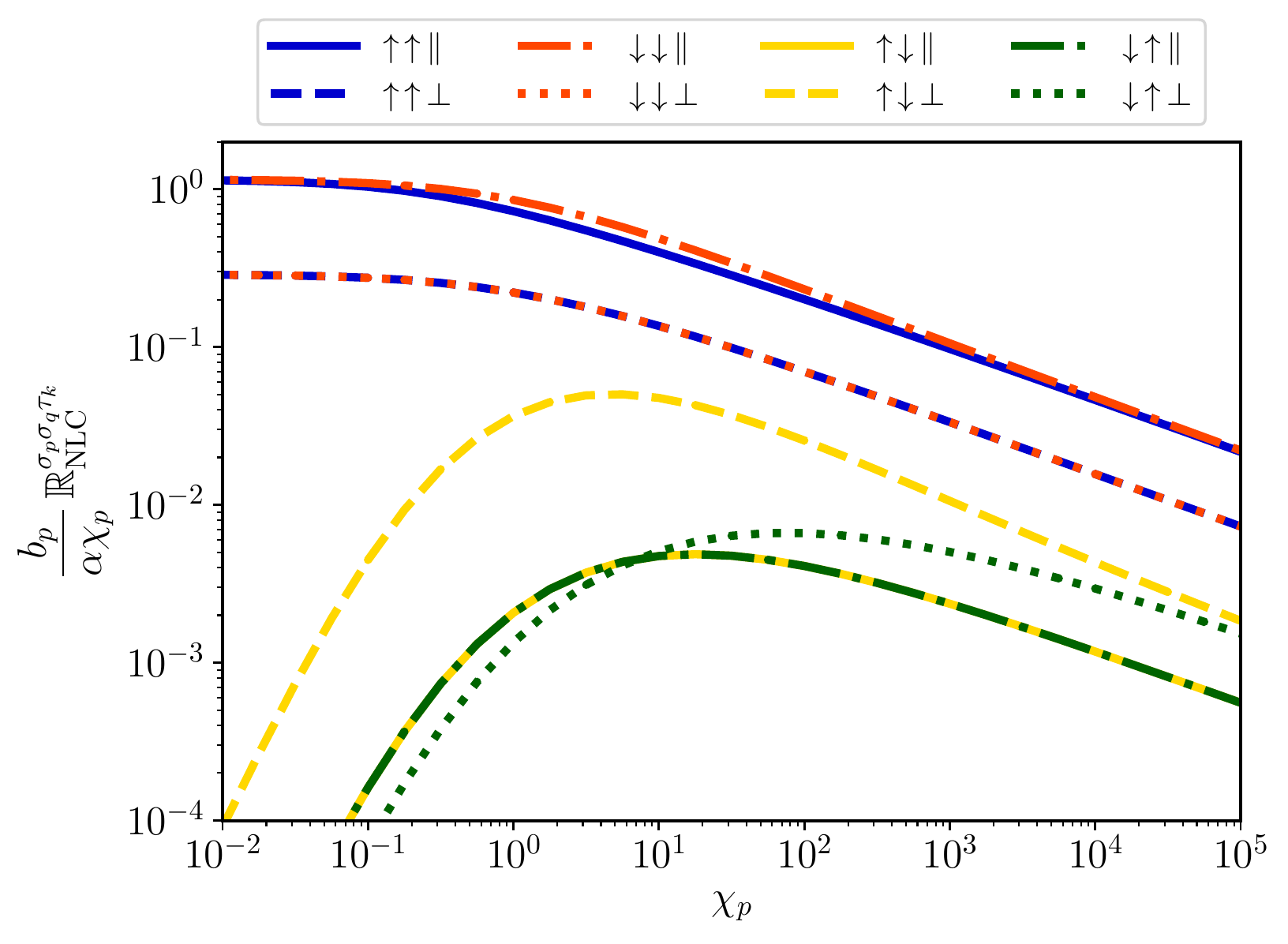}
    \caption{Polarisation resolved total rates for nonlinear Compton scattering as a function of the electron quantum parameter $\chip$.}
    \label{fig:PEtotal}
    \end{figure}

    \begin{figure}[!th]
    \centering
    \includegraphics[width=\columnwidth]{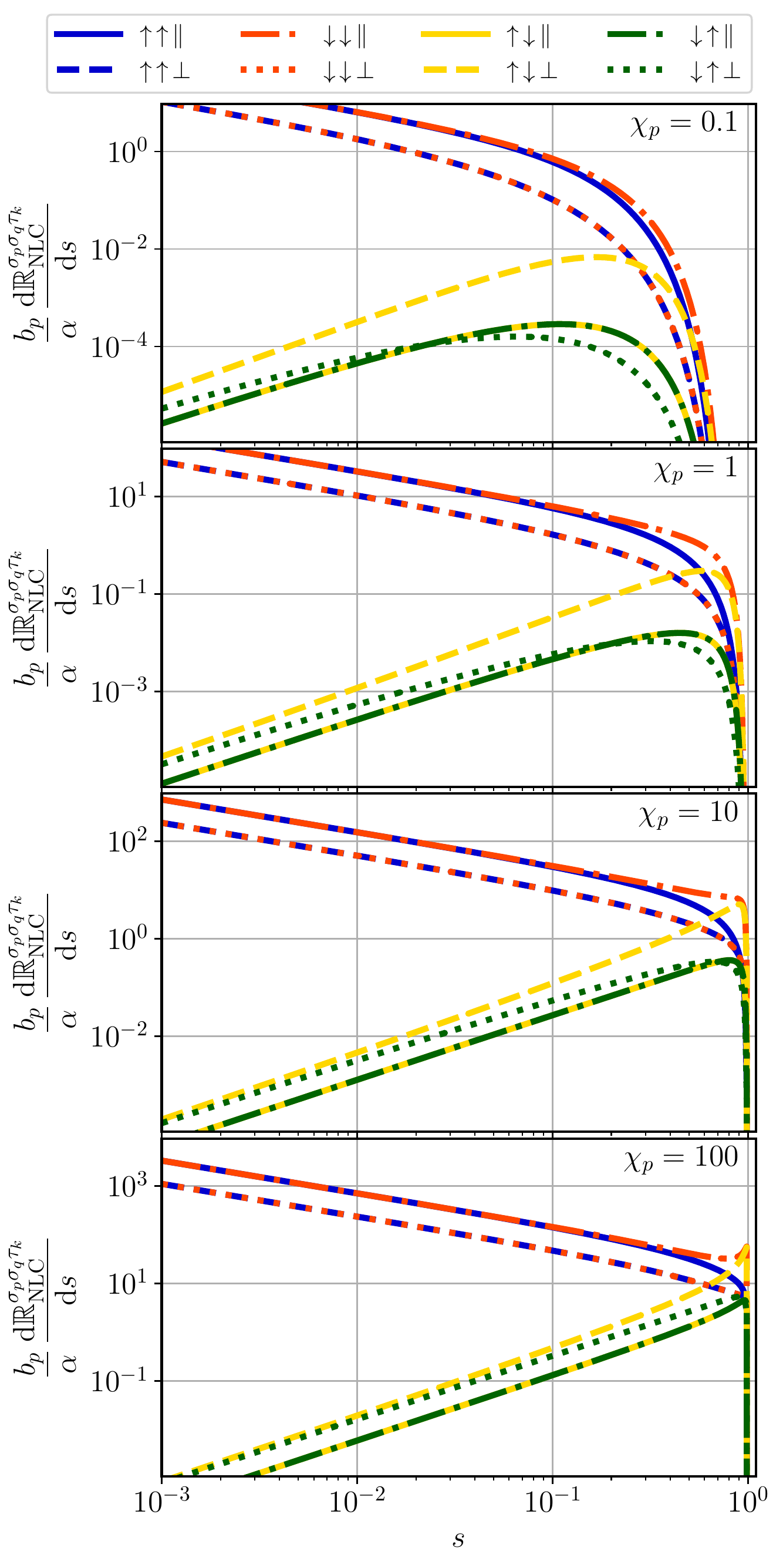}
    \caption{Plots of the polarisation resolved differential Compton spectra as functions of the normalised photon light-front momentum $s=k^-/p^-$ for four different values of $\chip$.}
    \label{fig:PE4panel_vert}
    \end{figure}

    In order to visualise how the differential photon spectrum comprises each polarisation channel, in \figref{fig:PE4panel_vert} we select four constant values of $\chip$ at different orders of magnitude: $\chip \in\{0.1,1,10,100\}$. In general, the hierarchy of the various polarisation channels can be different in the low-energy infra-red part of the spectrum compared to the high-energy UV part of the spectrum. For small $s\to 0$ (where $s$ is the fraction of photon light-front momentum) the rate of spin-flip channels go to zero, showing that the well known (integrable) infrared divergence of the polarisation averaged LCFA rates originates solely in the non-flip channels. For larger values of $s$ the non-flip and spin-flip channels approach each other, and eventually the hierarchy even changes with certain spin-flip channels becoming larger than some non-flip channels. (In other words, it is not just the flip of the spin that determines the hierarchy of rates.) As $\chip$ is increased, the part of the spectrum where the hierarchy between polarisation channels changes, moves to larger values of $s$. (Note that by the conservation law \eqnref{eq:emc-lf} the final state electron normalised light-front momentum is just $1-s$.)

    Also as $\chip$ increases, a new spectral feature develops in the high-energy part of the spectrum at $s\approx1$. In \figref{fig:PE1panel_linear} we show the development of this ``UV shoulder'' in more detail. Whilst the UV shoulder is known to exist and to develop into a pronounced peak approximately located at $s\sim1- 4/3\chip$ for $\chip \gg1$ \cite{bulanov13,tamburini19}, we see from Fig.~\ref{fig:PE1panel_linear} that only two of the eight polarisation channels are significantly contributing to it, with a strong correlation between the spin/polarisation states of all particles for this high-energy feature. This is particularly apparent in the right panel of Fig.~\ref{fig:PE1panel_linear}. For incident down electrons a $\parallel$-photon is emitted and the electron stays in a down state. For incident up electrons, a $\perp$-photon is emitted while the electron flips to a down state. Thus, by controlling the incident electron polarisation one could control the polarisation of the generated gamma rays in this high energy feature of the spectrum. Because the photons have very high energy, almost all of the incident electron energy is transferred to the photon. The existence of the UV shoulder can be clearly seen in calculations of two-step part of second-order processes such as nonlinear trident (NLC followed by NBW) \cite{king13b,king18,torgrimsson18,torgrimsson20}, and its existence has been commented on as contributing to free-particle ``shower'' type cascades \cite{king13a}.
 
 \begin{figure}[!bht]
    \centering
    \includegraphics[width=\columnwidth]{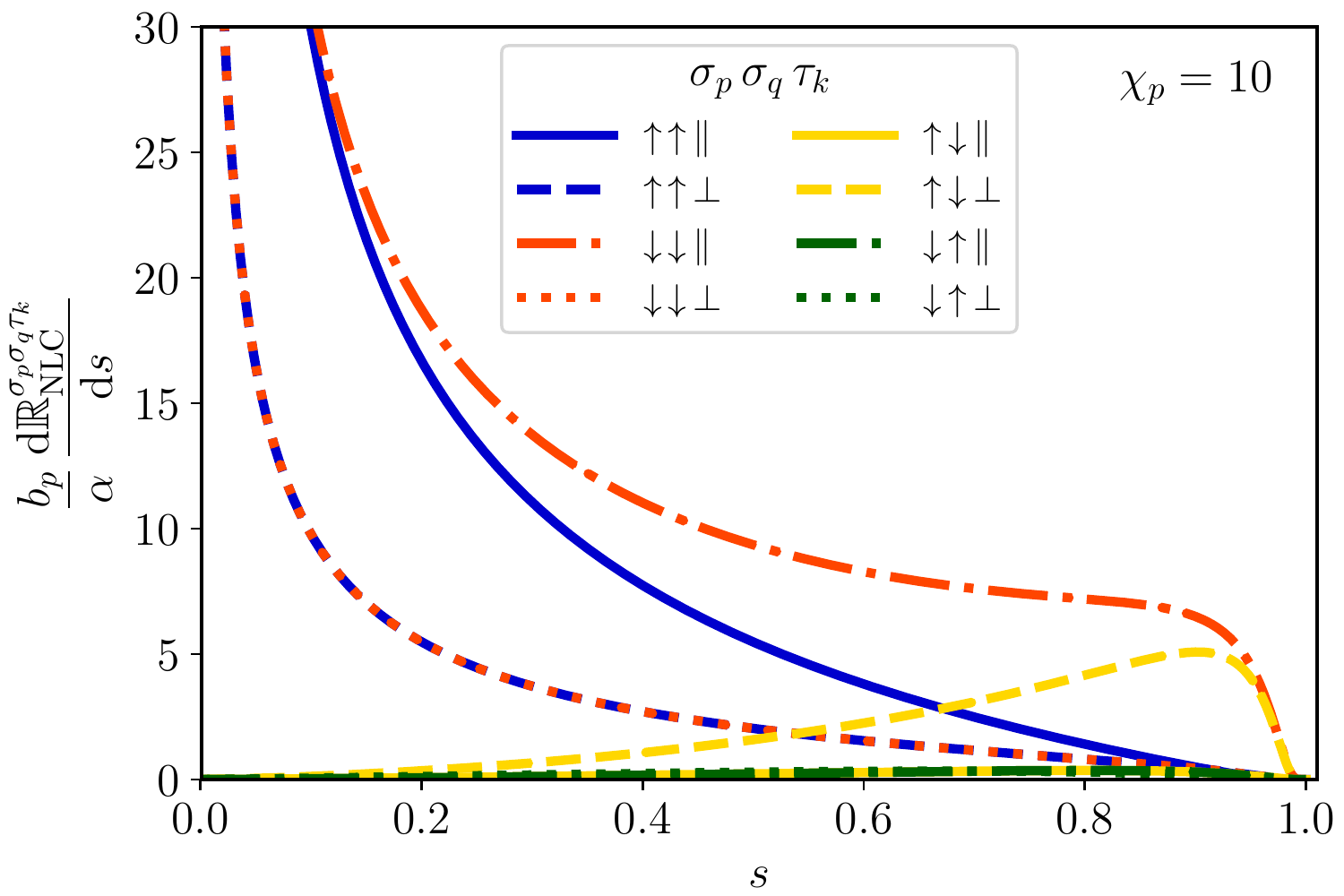}
    \includegraphics[width=\columnwidth]{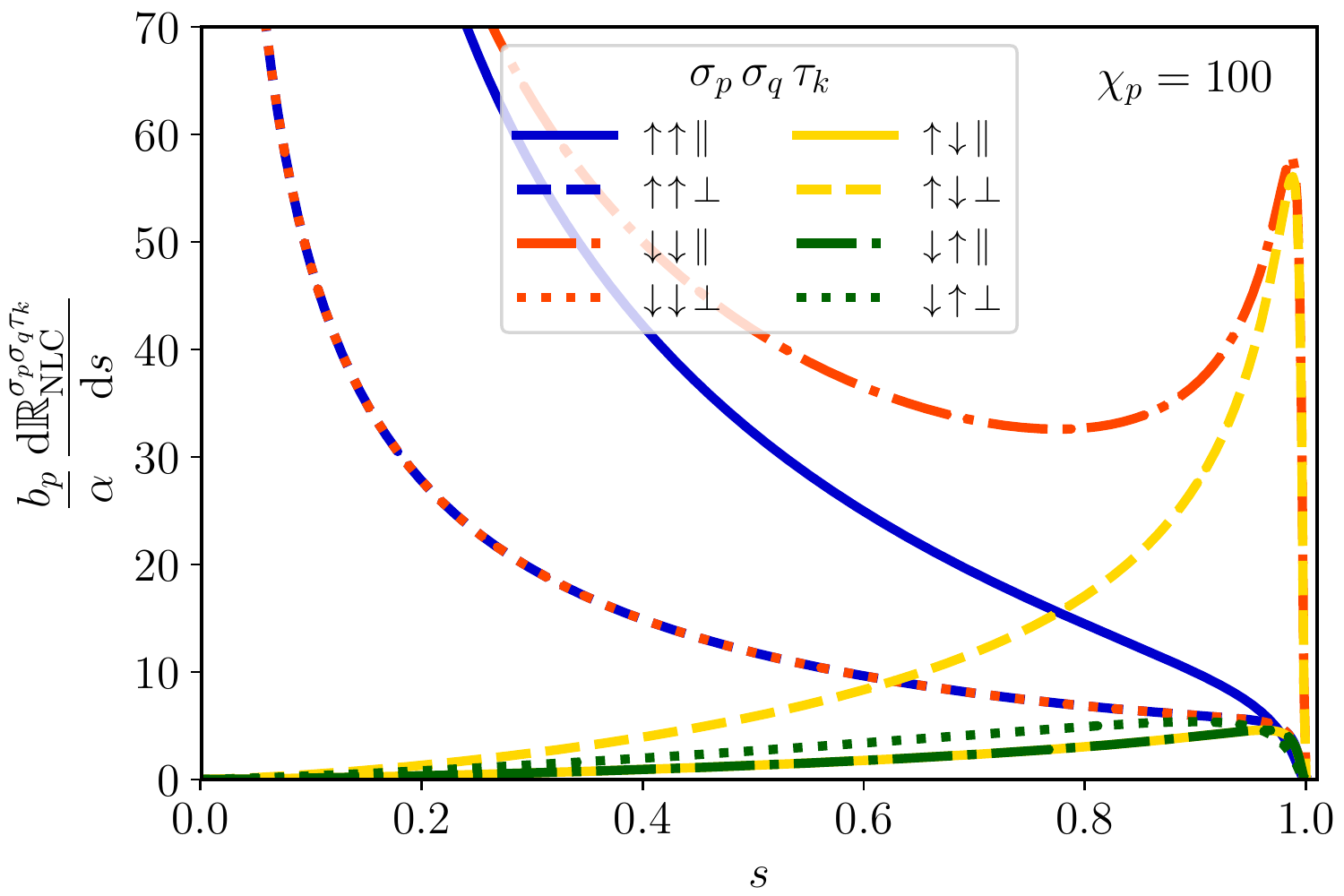}
    \caption{Differential Compton rate for $\chip = 10$, $100$ on a linear scale, highlighting the formation of the UV shoulder at $s\simeq 1$ for large $\chip$ which has a very strong polarisation dependence. The electrons emerging from this interaction are strongly down-polarised. There is also a strong correlation between the initial electron polarisation and the polarisation of the emitted photons in the shoulder.
    }
    \label{fig:PE1panel_linear}
    \end{figure}

    Although in this section we have thus far focussed on the NLC process for \emph{electrons}, analogous arguments apply to the NLC process for \emph{positrons}. We note here the necessary changes. First, in the classical kinetic momentum of the electron in a plane wave $\pi_p(\phi)$ from \eqnref{eqn:pi1}, $a=eA$, where $e<0$ for an electron. The charge of the positron is positive, $|e|$. Thus, the classical kinetic momentum of a positron differs from that of an electron. The correct expression taking into account the change in the sign of the charge, is given by $-\pi_{-p}(\phi)$.
 
    Moreover, the different sign of the charge for electrons and positrons implies that the vector of the magnetic moment and spin are parallel in one case and antiparallel in the other case. That means, the spin-field interaction has the opposite sign for positrons. Thus, in order to employ the electron NLC rates, \eqnref{eq:RNLCfinal}, for positrons one also has to make the replacements: $\sigma_{p} \to - \sigma_{p}$ and $\sigma_{q} \to - \sigma_{q}$. It is evident that this affects neither the terms in \eqref{eq:RNLCfinal} containing the product $\sigma_p \sigma_q$ nor does it affect the spin-averaged rates, Eqns.~\eqref{eq:NLCspinunpol} and \eqref{eq:NLCunpol}.

    \subsection{Asymptotic Limits}

    For the discussion of the asymptotics of the rate for large and small values of $\chip$ it is convenient to treat spin flip ($\sigma_q = -\sigma_p$) and non-flip ($\sigma_q=\sigma_p$) separately, as we will find them to have different asymptotic behaviour. 
    {Here we choose the quantum parameters, $\chi_e =
    \chip$ and $\chi_\gamma=\chik$, occurring in the LCFA, to take constant values, which is equivalent to considering the case of a constant crossed field background.}

    \subsubsection{$\chip \ll 1$}

    For nonlinear Compton scattering, the $\chik$ parameter of the emitted photon (which is bounded above by the $\chip$ parameter of the initial electron) quantifies the recoil when the electron emits a photon. Furthermore, the incoming electron parameter, $\chip$, is such that $\chip,\chik \propto \hbar$. Therefore, the limit of $\chip\to0$ is synonymous with the classical limit. The asymptotic expansion of the total NLC rate $\mathbb R_\mathrm{NLC}$ for small $\chip \ll 1$ can be derived by changing the integration variable from light-front momentum fraction $s$ to $z$ (the argument of the Airy functions) and performing a systematic power series expansion in $\chip$, yielding
    \begin{align}
        \mathbb R^{\sigma_p,\sigma_p,\tau_k}_\mathrm{NLC}
            &\sim \frac{\alpha \chip }{b_p} \frac{1}{2\sqrt{3}} 
            \left[
            \frac{5}{2}+ \frac{3}{2}\tau_k \right. \nonumber \\
            &
            \quad -  \left(\frac{3}{4} \sigma_p  (1 + \tau_k)+\frac{ 4 + 3 \tau_k}{\sqrt{3}}\right)   \chip   
              \nonumber \\
            & \left. \quad
        +\left(\frac{5}{2} \sqrt{3} \sigma_p  (1 + \tau_k )+\frac{5}{48} (75 + 62 \tau_k)\right) \chip ^2 
        \right]\,, \label{eqn:asysmall1}\\
        \mathbb R^{\sigma_p,-\sigma_p,\tau_k}_\mathrm{NLC} 
            &\sim \frac{\alpha \chip^3 }{b_p} \frac{1}{2\sqrt{3}} 
            \left[
                \frac{15}{16} - \frac{5}{6} \tau_k 
                + \frac{\sqrt{3}}{2} \sigma_p ( 1 - \tau_k )
            \right] \,, \label{eqn:asysmall2}
    \end{align}
    as $\chip\to0$.
    In the non-flip rate, \eqnref{eqn:asysmall1}, the leading order is $\mathcal O(\chip)$, and the leading order is independent of the spin state of the incoming electron. A spin-splitting (difference between up and down incident electrons) only occurs in the order $\mathcal O (\chip^2)$ and only for $\parallel$ photon polarisation, $\tau_k = +1$ (there is no spin-splitting at all for the $\perp$ polarisation). For the spin-flip rate, the leading term is suppressed at $\mathcal O(\chip^3)$. Here, the leading term does show spin-splitting, but only for the $\perp$ photon polarisation ($\tau_k=-1$). The overall leading order of the photon emission rate agrees with the classical radiation. The suppression of spin-effects in the NLC rates for small $\chip \ll 1$ is consistent with the fact that spin is a quantum property and spin-sensitive effects should disappear in the classical limit.

    \subsubsection{$\chip \gg 1$}

    The asymptotic expansion of the NLC rate for large $\chip \gg 1$ can be calculated by first perturbatively expanding the Airy functions for small argument $z$. The resulting integrals can be easily performed for the leading order terms stemming from the $\Ai$ and $\Ai'$--terms, yielding
    \begin{align}
        \mathbb R^{\sigma_p,\sigma_p,\tau_k}_\mathrm{NLC} 
                 &\sim \frac{\alpha\chip^{2/3}}{b_p} \frac{\Gamma(\frac{2}{3}) }{ 18 \cdot 3^{1/3}}
                 \left[ \vphantom{\frac{1}{1}}
                         13 \left( 1 + \frac{\tau_k}{2}\right)
                        \right. \nonumber \\
                & \qquad \left.
                         - (3\chip)^{-1/3}  \,  \sigma_p (1 + \tau_k)
                           \frac{ 7 }{ 2 } \frac{ \Gamma(\frac{1}{3} ) }{\Gamma(\frac{2}{3})}
                 \right] \,, \label{eqn:asybig1}\\
        \mathbb R^{\sigma_p,-\sigma_p,\tau_k}_\mathrm{NLC} 
                 &\sim \frac{\alpha\chip^{2/3}}{b_p} \frac{\Gamma(\frac{2}{3}) }{ 18 \cdot 3^{1/3}}
                 \left[ \vphantom{\frac{1}{1}}
                 1 - \frac{\tau_k}{2} \right. \nonumber \\
                & \qquad \left.
                 + (3\chip)^{-1/3} \,  \sigma_p (1 - \tau_k)
               \frac{ 5 }{ 2 } \frac{ \Gamma(\frac{1}{3} ) }{\Gamma(\frac{2}{3})  }
                 \right] \,, \label{eqn:asybig2}
    \end{align}
    as $\chip\to\infty$.
    In this asymptotic limit, for both the spin-flip and non-flip rates the leading order term is $\mathcal O(\chip^{2/3})$ and independent of the spin of the incident electron. Spin dependence only occurs in the next to leading order, which is $\mathcal O(\chip^{1/3})$. This term completely vanishes for unpolarised electrons, where the next non-vanishing term is $\mathcal O(1)$.

    \begin{figure}[h!!]
	\centering
	\includegraphics[width=\columnwidth]{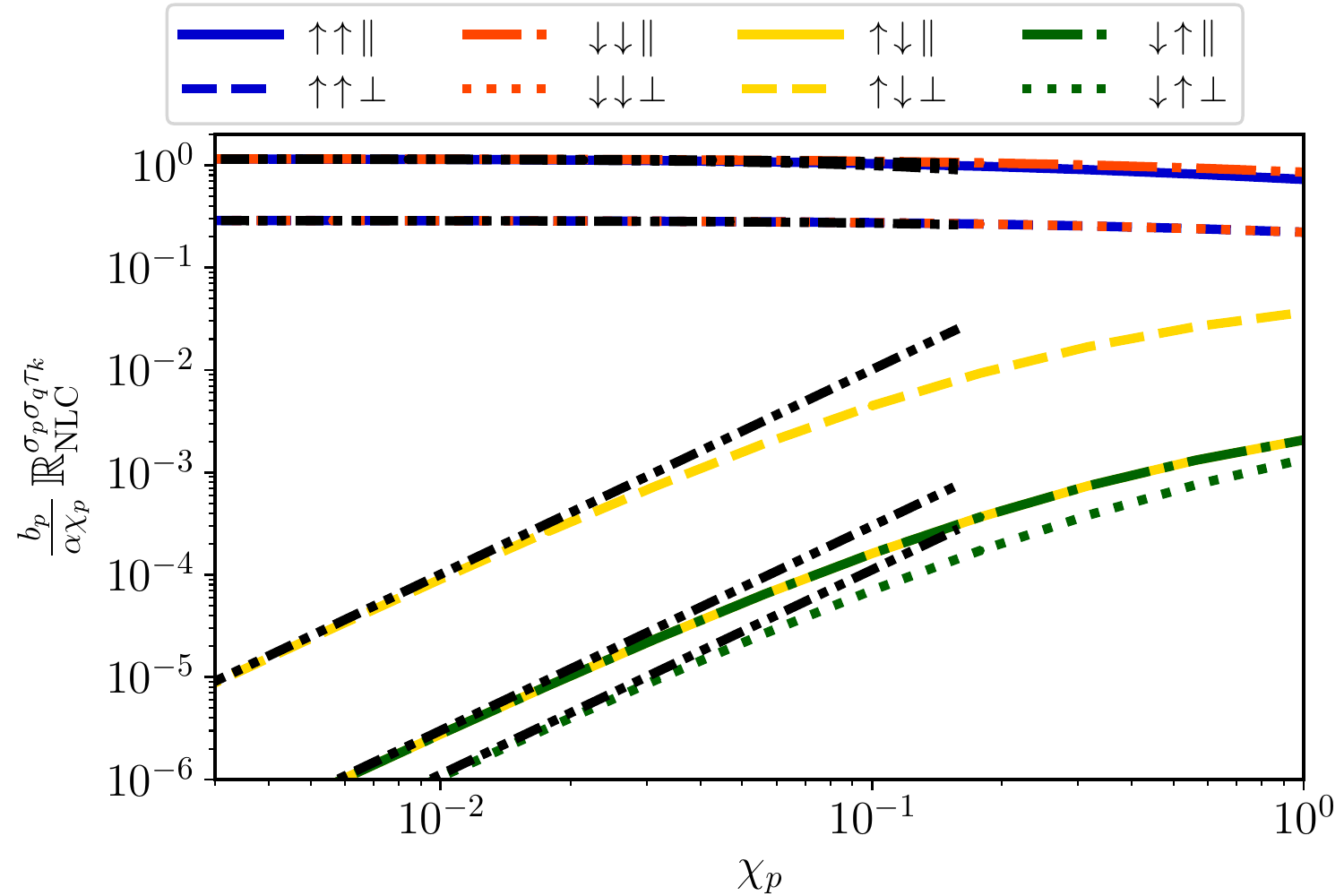}
	\includegraphics[width=\columnwidth]{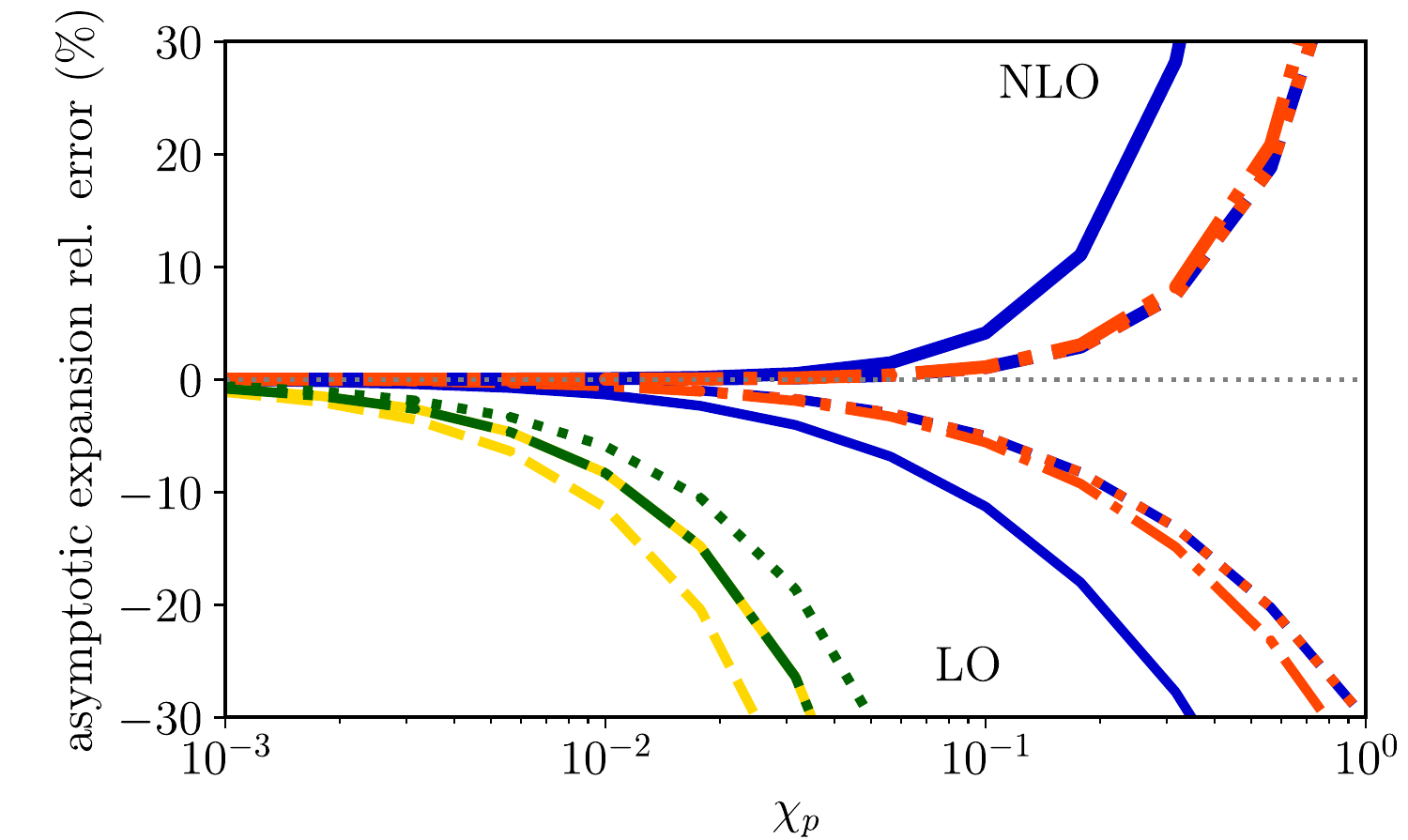}
	\caption{Comparison of the total nonlinear Compton rates (colored curves) with their asymptotic expansions (black dash-dotted curves) for $\chip \ll 1$ (top) and relative error of the asymptotic expansion (bottom). We compare the leading order (LO) and the next-to-leading order (NLO) for the non-flip rates.}
	\label{fig:PEtotal_asy_small}
	\end{figure}

    \begin{figure}[h!!]
	\centering
	\includegraphics[width=\columnwidth]{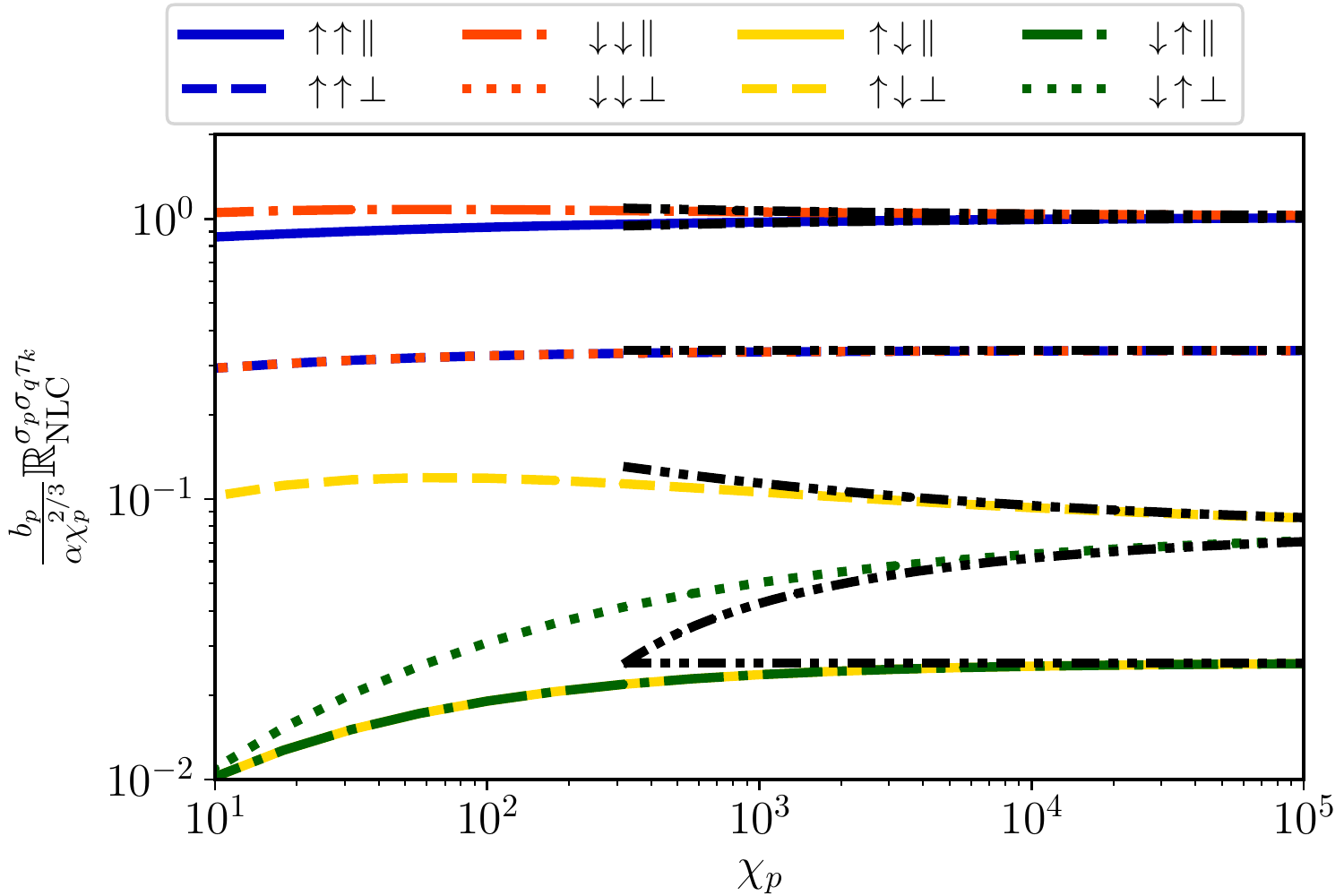}
	\includegraphics[width=\columnwidth]{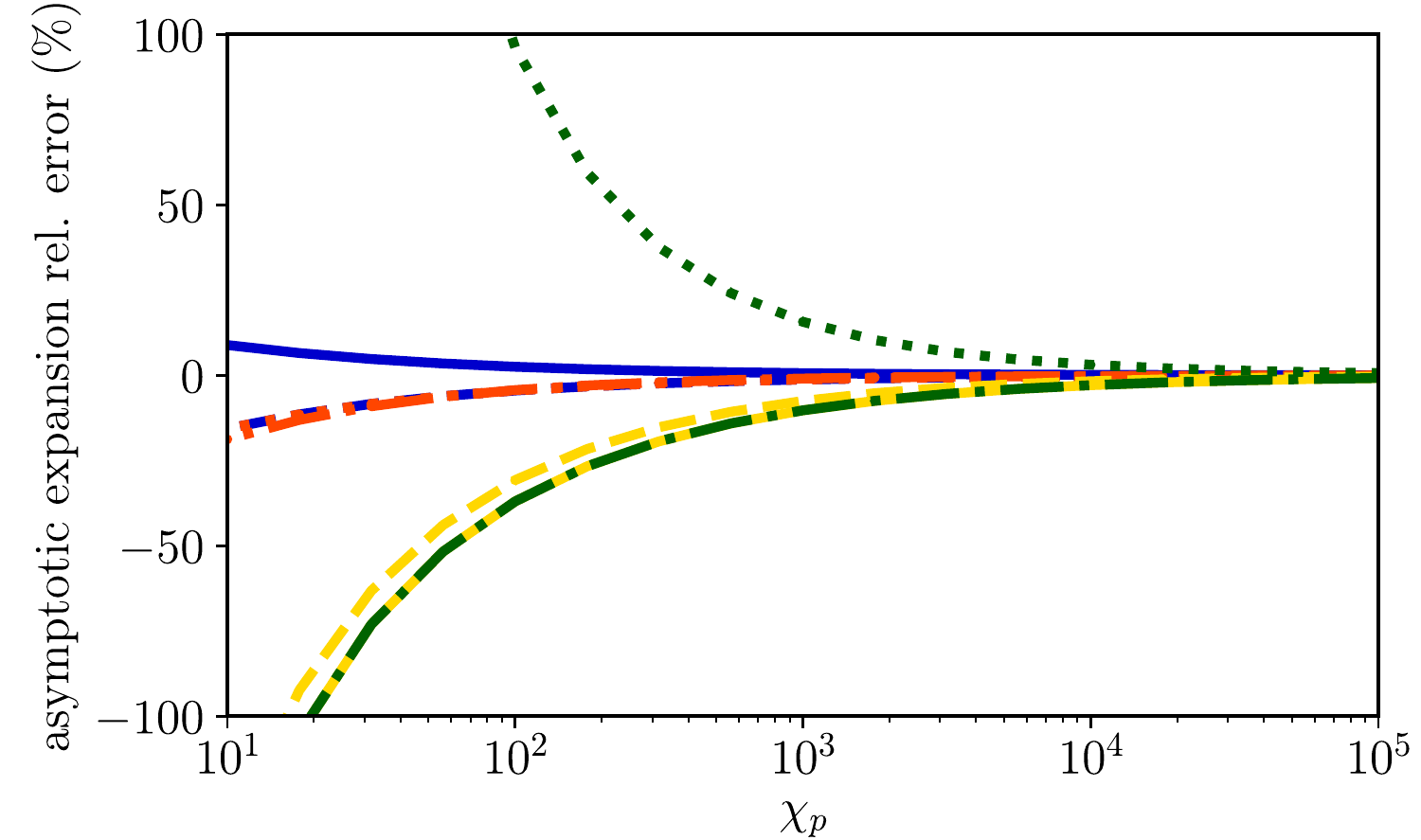}
	\caption{
		Comparison of the (scaled) total nonlinear Compton rates with their asymptotic expansions (black dash-dotted curves) for $\chip \gg 1$ (top) and relative error of the asymptotic expansion (bottom).
	}
	\label{fig:PEtotal_asy_large}
	\end{figure}

    To illustrate the asymptotic scaling of the relations in Eqs.~\eqref{eqn:asysmall1}--\eqref{eqn:asybig2}, and their accuracy, the $\chip \ll 1 $ and $\chip \gg 1$ parts of the total yield have been highlighted in \figref{fig:PEtotal_asy_small} and \ref{fig:PEtotal_asy_large}, respectively. As already commented above, we see that in the $\chip \to 0$ limit, all spin-flip channels are suppressed by a factor $\chip^2$ compared to the non-flip channels. However, we also see that the value of $\chip$ at which the asymptotic scaling reaches a prescribed level of accuracy, changes, depending on the order of the scaling. To make this manifest, in \figref{fig:PEtotal_asy_small} (bottom) we plot the relative error of the asymptotic expression, as a function of $\chip$. Generally speaking, to arrive at a given accuracy, the asymptotic relations for the spin-flip channels require $\chip$ to be an order of magnitude more asymptotic, e.g. in the case $\chip\ll1$, an order of magnitude smaller than for the non-flip channels. For example, in the $\chip\to 0$ limit, a $10\%$ accuracy is reached by the non-flip relations already at $\chip \approx 0.1$, whereas it requires $\chip \approx 0.01$ for the same accuracy in the asymptotic relations of the spin-flip channels. Likewise, it is remarkable that the asymptotic expressions in the $\chip\to \infty$ limit, see Fig.~\ref{fig:PEtotal_asy_large} (bottom), only reach an accuracy of $10\%$ for $\chip \gtrsim 10^3$ for the spin-flip channels. For the spin-flip channel and emission of photon into the $\perp$ polarisation, this accuracy is reached at an order of magnitude even larger than this. With present day laser and accelerator technology one can only reach values of $\chip \lesssim 10$, and so large $\chi_p$ asymptotic expressions can only be used cautiously.

    \section{Nonlinear Breit-Wheeler Pair Production}

    \subsection{S-matrix}
    To calculate the probability for nonlinear Breit-Wheeler pair production we need to utilise the Volkov state for an (outgoing) positron, which is given by \cite{dipiazza18,muller11}
    \begin{align} \label{eq:volkov-positron}
         \Psi^{(-)}_{p\sigma_p}(x)    = E_{-p}(x)  \, v_{p\sigma_p} \,,
    \end{align}
    with the Ritus matrices, \eqnref{eq:ritusmat}, constant positron bi-spinors $v_{p \sigma_p}$, and
    where the superscipt ``$-$'' signifies that the positron Volkov state is a negative energy solution of the Dirac equation \eqref{eq:dirac}. Employing \eqref{eq:volkov-positron}, the S-matrix element of NBW, see Fig.~\ref{fig:feynman} (right), can be expressed as follows:
    \begin{multline}
    S_\mathrm{NBW}(kj\to p\sigma_p;q\sigma_q) 
    	\\ 
    	= \int \! \ud^4 x \: \bar \Psi_{q\sigma_q}(x) [-ie \slashed \epsilon_j e^{-ik.x} ] \Psi^{(-)}_{p\sigma_p}(x)  
    	\\
    	= -ie (2\pi)^4 \int \frac{\ud \ell}{2\pi} \delta^{(4)}(k+\ell \kappa - q - p ) \: \mathscr M_\mathrm{NBW}
    \end{multline}
    We emphasise that $p$ ($\sigma_p$) is the four-momentum (spin index) of the created positron and $q$ ($\sigma_q$) refers to the electron. The nonlinear Breit-Wheeler amplitude 
    \begin{multline}
    \mathscr M_\mathrm{NBW}(\sigma_p,\sigma_q,j) = \\
    \Lambda_{\mu,j}
    \int \! \ud \phi \: e^{-i \int \frac{k.\pi_{-p}}{\kappa.q } \ud \phi } \: \bar u_{q\sigma_q}  \mathscr J_\mathrm{NBW}^\mu (\phi) v_{p\sigma_p}
    \end{multline}
    can be expressed in terms of the current
    \begin{multline} \label{eq:current.NBW}
    	\mathscr J^\mu_\mathrm{NBW}(\phi) = \gamma^\mu 
    	+ \left[ \frac{\slashed a \slashed \kappa \gamma^\mu }{2(\kappa.q)} - \frac{ \gamma^\mu \slashed \kappa \slashed a}{2(\kappa. p)} \right] h(\phi) \\
     	- \frac{\slashed a \slashed \kappa \gamma^\mu \slashed \kappa \slashed a}{4(\kappa.p)(\kappa.q)} h^2(\phi) \,,
    \end{multline}
    where the kinetic momentum of the positron is given by $-\pi_{-p} (\phi)$, with $\pi_p(\phi)$ given in Eq.~\eqref{eqn:pi1}.
    The Dirac trace for NBW is:
    \begin{multline}
    \mathsf T^{\mu\nu} 
    =  \frac{1}{4} \mathrm{tr}\: \Big[
    	(\slashed q + m ) (1+\sigma_q \gamma^5 \slashed \zeta_q) \mathscr J^\mu(\phi) \\
    	\times 
    	(\slashed p - m ) (1+\sigma_p \gamma^5 \slashed \zeta_p) \bar{  \mathscr J}^\nu(\phi ' )
        \Big] 
    \end{multline}
    and using the current from \eqref{eq:current.NBW},
    can be decomposed into four parts: unpolarised ($\UP$), electron polarised ($\EP$), positron polarised ($\PP$), and polarisation correlation ($\PC$), which are defined as follows:
    \begin{align}
     \mathsf T^{\mu\nu} (\sigma_p,\sigma_q) & = 
    		 \UP^{\mu\nu} 
    		+ \sigma_q \EP^{\mu\nu} 
    		+ \sigma_p \PP^{\mu\nu} 
    		+ \sigma_p \sigma_q \PC^{\mu\nu} \,,
    \end{align}
    with the four contributions
    \begin{align} \label{eq:NBWtraceUP}
    \UP^{\mu\nu} & \equiv
    \frac{1}{4} \TR{ (\slashed q + m ) \: \mathscr J^\mu(\phi) \: (\slashed p - m) \: \bar{ \mathscr J }^\nu(\phi') 
            \vphantom{\slashed \zeta_q} }  \,, \\
     \EP^{\mu\nu} & \equiv
    \frac{1}{4} \TR{  (\slashed q + m ) \: \gamma^5\slashed \zeta_q \: \mathscr J^\mu(\phi)  \: (\slashed p - m ) \: \bar{ \mathscr J }^\nu(\phi') } \,,  \\
     \PP^{\mu\nu}  & \equiv
    \frac{1}{4} \TR{  (\slashed q + m ) \: \mathscr J^\mu(\phi) \:  (\slashed p - m )\: \gamma^5\slashed \zeta_p \: \bar{ \mathscr J }^\nu(\phi') } \,, \\
     \PC^{\mu\nu} & \equiv
    \frac{1}{4} \TR{   (\slashed q + m ) \: \gamma^5\slashed \zeta_q \: \mathscr J^\mu(\phi) \: 
    	(\slashed p - m ) \: \gamma^5\slashed \zeta_p \: \bar{ \mathscr J }^\nu(\phi') } \,.
    \label{eq:NBWtracePC}
    \end{align}

    \subsection{Pair Production Probability}

    The evaluation of the traces for NBW pair production is presented in detail in Appendix \ref{app:pp}. With those, and after performing the integration over the transverse momentum of the outgoing positron, we find the fully polarisation resolved NBW pair production probability in a linearly polarised plane wave of arbitrary shape:
    \begin{widetext}
    \begin{align}
    \frac{\ud \mathbb P_\mathrm{NBW}}{\ud s } & = \frac{\alpha}{8\pi b_k} 
            \int \! \ud \varphi  \int_{-\infty}^\infty \! \frac{\ud \theta}{-i\theta} 
            \:e^{i\theta \mu \tilde x_0} \: \mathcal I_\mathrm{NBW} \,, \\
    \mathcal I_\mathrm{NBW} &= 1 + \sigma_p \sigma_q +\tau_k \sigma_p \sigma_q (1-\tilde g)
                             + \frac{\xi^2 \theta^2 \langle \dot h \rangle^2}{2} \left[ \tilde g + \sigma_p \sigma_q \right]
                             - \xi^2 \Delta h^2  \, \tau_k \left[ 1 + \tilde g \sigma_p \sigma_q \right] \nonumber \\
                & \qquad    - \frac{i\theta \xi \langle \dot h \rangle }{2}
                    \left[
                    \frac{\sigma_p}{s} - \frac{\sigma_q}{1-s} 
                    +\tau_k \left( \frac{\sigma_q}{s} - \frac{\sigma_p}{1-s}
                    \right)
                    \right] \,,
    \end{align}
    \end{widetext}
    with the positron's light-front momentum fraction $s = p^-/k^-$, $\tilde g = 1 - \frac{1}{2s(1-s)}$, Kibble mass \eqnref{eq:Kibblemass} and $\tilde x_0$ defined in \eqnref{eq:tildex0}. The initial photon is in a polarisation state $\epsilon$ characterised by the Stokes parameter $\tau_k = |c_1|^2-|c_2|^2$, where $\epsilon = c_1 \Lambda_1 + c_2 \Lambda_2$. In addition, the definition of $\Delta h^2$ given below Eq.~\eqref{eqn:pulse1}, as well as the statements about regularization apply here as well.

    Details of the derivation of the LCFA, including the integrals over the phase variable $\theta$ are collected in Appendix \ref{app:pp}. Here we present the final result for the completely polarised NBW pair production rate within the LCFA
    \begin{widetext}
    \begin{align}
    \frac{\ud \mathbb R_\mathrm{NBW}}{\ud s} (  \sigma_p , \sigma_{q} , \tau_k ) 
    &=
     \frac{\alpha}{4b_k}
     \left[
      \{ 1 + \sigma_p \sigma_q + \tau_k \sigma_p\sigma_q (1 - \tilde g)   \} \: \Ai_1(\tilde z)
     \vphantom{\frac{1}{1}}\right. 
     \nonumber \\
     & \qquad \quad +	
     \left\{ \frac{\sigma_p}{s} - \frac{\sigma_q}{1-s} 
     + \tau_k
     \left( \frac{\sigma_q}{s} - \frac{\sigma_p}{1-s} \right) 
     \right\}
     \frac{\Ai(\tilde z)}{\sqrt{\tilde z}} \: \shd
     \nonumber \\
     & \qquad \quad +  \left. 
     \left\{ ( \tilde g + \sigma_p\sigma_q ) + \tau_k \frac{1+\tilde g \sigma_p \sigma_q}{2} \right\} \frac{2\Ai'(\tilde z)}{\tilde z} 
     \right] \label{eq:rate.pp.final.stokes}
    \end{align}
    \end{widetext}
    where the argument of the Airy functions is given by $\tilde z = \left[\chi_\gamma(\varphi) s(1-s)\right]^{-2/3}$.
    The photon quantum parameter $\chi_\gamma(\varphi)$ again refers to the local value in the field, given by
    $\chi_\gamma(\varphi) = \chi_k |\dot h(\varphi)|$, 
    where $\chi_k = \xi b_k$. The quantum energy parameter $b_k = k.\kappa/m^2$ is related to the center-of-mass energy of the incident photon colliding with the plane wave laser field. We emphasise again that $s = p.\kappa / k.\kappa$ is the fractional light-front momentum of the \emph{positron} in relation to the light-front momentum of the incident photon. Likewise, $\sigma_p$ refers to the spin state of the positron, and $\sigma_q$ to the spin-state of the electron.

    It is straightforward to recover expressions for totally or partially unpolarised channels. For instance, for the decay of a polarised photon into an unpolarised pair we have to sum over all fermion polarisations, which is equivalent to setting $\sigma_p = \sigma_q = 0$ and multiplying the result by 4:
    \begin{multline} \label{eq:NBW.photpol}
    4\, \frac{\ud \mathbb R_\mathrm{NBW}}{\ud s} (  \sigma_p=0 , \sigma_q=0,\tau_k ) 
    \\ 
    =
    \frac{\alpha}{b_k}
    \left[  \Ai_1(\tilde z)
     +	
    \{ 2  \tilde g  +   \tau_k   \} \frac{\Ai'(\tilde z)}{\tilde z} 
    \right] \,.
    \end{multline}
    This agrees with expressions from the literature \cite{king13a} (Sometimes in the literature the Stokes parameter is expressed as $\tau_k = \cos 2\vartheta$, where $\vartheta$ is the angle of the photon polarisation in relation to the laser polarisation, characterised by $\Lambda_1$).

    To obtain the completely unpolarised NBW rate we have to average \eqnref{eq:NBW.photpol} over the incoming photon polarisation by setting $\tau_k=0$:
    \begin{multline}
    4\, \frac{\ud \mathbb R_\mathrm{NBW}}{\ud s} (  \sigma_p=0 , \sigma_q=0,\tau_k=0 ) 
    \\ =
    \frac{\alpha}{b_k}
    \left[  \Ai_1(\tilde z) +	2  \tilde g  \frac{\Ai'(\tilde z)}{\tilde z} 
    \right] \,.
    \end{multline}

    We can also find the result for the production of a polarised pair by unpolarised photons by just setting $\tau_k=0$
    \begin{multline}
        \frac{\ud \mathbb R_\mathrm{NBW}}{\ud s} (  \sigma_p , \sigma_{q} , \tau_k=0 ) 
        = \\
        \frac{\alpha}{4b_k}
        \left[
        \{ 1 + \sigma_p \sigma_q   \} \: \Ai_1(\tilde z)
        \vphantom{\frac{1}{1}}  +	
        \{ 2 ( \tilde g + \sigma_p\sigma_q )  \} \frac{\Ai'(\tilde z)}{\tilde z} \right.
        \\
        + \left.
        \left\{ \frac{\sigma_p}{s} - \frac{\sigma_q}{1-s} 
        \right\}
        \frac{\Ai(\tilde z)}{\sqrt{\tilde z}}\: \shd 
        \right] \,.
    \end{multline}

    \subsection{Discussion of the Pair Production Rates}

    \begin{figure}[h!!]
    \centering
    \includegraphics[width=\columnwidth]{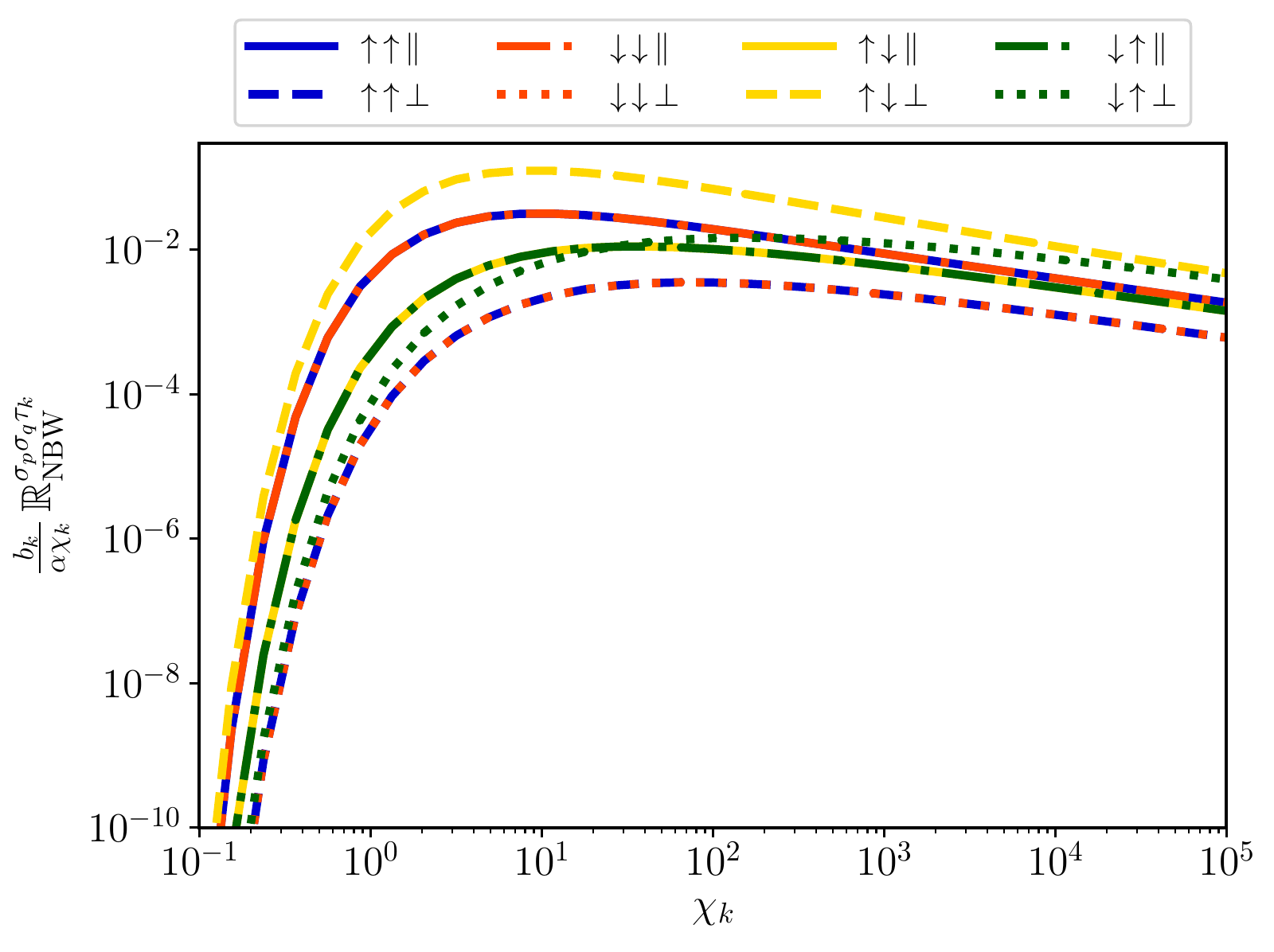}
    \caption{Scaled total spin/polarisation resolved NBW pair production rates as a function of $\chik$.}
    \label{fig:PPtotal}
    \end{figure}

    We illustrate the results for the various polarisation channels of NBW pair production in a series of plots, starting with the total rates in \figref{fig:PPtotal}. All of the eight channels are strongly suppressed for small $\chik$. This is a reflection of the fact that NBW pair production, unlike NLC, is a pure quantum process that must vanish in the classical limit as $\chik \to 0$. (See also the detailed discussion of the asymptotic behaviour below.) Similar to NLC scattering, the plot of total NBW rates (see \figref{fig:PPtotal}) shows a certain hierarchy of the polarisation channels which does not change with $\chik$, apart from one particular channel where a $\perp$-photon produces a pair with positron spin $\sigma_p=\downarrow$ and electron spin $\sigma_q=\uparrow$. In this channel the pair is produced in its least favourable spin state since the electron spin is aligned parallel to the magnetic field and the positron is aligned anti-parallel to the magnetic field. This channel is one of the smallest contributions to the overall rate for small $\chik \ll1$, but is one of the dominant ones for $\chik \gg1$.
    In general, the most probable channel is the one in which a photon polarised in the $\perp$ state decays into a pair in which the spins are aligned such that their interaction energy with the background magnetic field is minimised, i.e.~the electron (positron) is aligned antiparallel (parallel) with the field \cite{ternov95}. This can be seen from the energy in the rest frame of the particle \cite{griffiths}, $ U_B = -\vec \mu \cdot \vec B$, with $\vec \mu = e g_e \vec s / 2m$ (and recalling $e<0$ for an electron), where $\vec s$ are the spatial components of the spin four-vector $S=\sigma_p\zeta_p$ in the electron rest frame.

    \begin{figure}[!bth]
    \centering
    \includegraphics[width=\columnwidth]{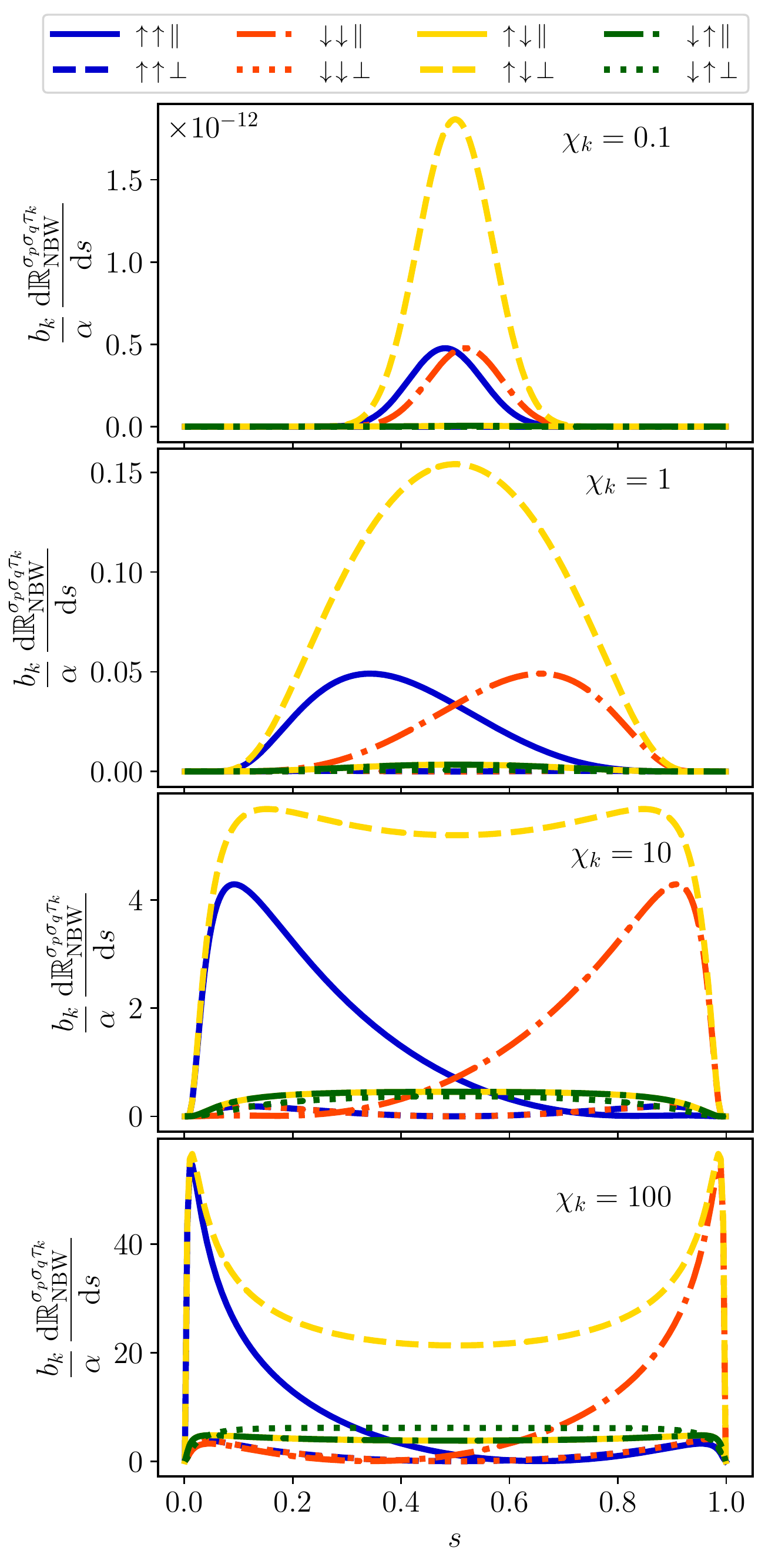}
    
    \caption{Spin/polarisation resolved differential NBW pair production rates. Observe how photons of different polarisation produce pairs with different symmetry properties. For the largest contribution which comes from $\perp$-photons the spectra are symmetric around $s=1/2$, and the pair has anti-parallel spins with both particles in the desired (lower energy) state. For $\parallel$-photons the pair has preferably parallel spins, with one high-energy particle and one low-energy particle being produced.}
    \label{fig:PP1panel}
    \end{figure}

    In Fig.~\ref{fig:PP1panel} we plot the light-front momentum spectrum of produced positrons for a range of incoming photon quantum parameters, $\chik \in\{0.1,1,10,100\}$. It is evident that, especially for smaller values of $\chik$, only a few channels are dominant. For increasing $\chik$ we see new peak structures appear close to $s\sim0,1$ which are strongly related to the ``UV shoulder'' seen in \figref{fig:PE1panel_linear} for NLC \cite{bulanov13}. The unpolarised pair production spectrum is symmetric around $s=1/2$, i.e. symmetric in the exchange of electron and positron $s\leftrightarrow 1-s$.  In Fig.~\ref{fig:PP1panel} we clearly see that not all polarisation resolved channels adhere to this symmetry. In particular the channels in which a $\parallel$-photon decays into a pair with parallel spins (i.e.~only one of the particles is in its preferred energy state), break the symmetry about $s=1/2$, meaning that one of the particles is preferably created with a higher momentum than the other one.

    \subsection{Asymptotic Limits}

    Here we provide and discuss the asymptotic limits for small and large constant values of $\chik$ for the spin and polarisation dependent pair production rates. Here it is convenient to distinguish the case of parallel spins $\sigma_q = \sigma_p$, and anti-parallel spins $\sigma_q = -\sigma_p$.

    \subsubsection{$\chik \ll 1$}
    
    For small $\chik$, the asymptotic scaling of the total NBW rate can be calculated by using the fact that for $\chik \ll 1$ the argument of the Airy functions $\tilde z$ is always large. Performing an asymptotic expansion of the Airy functions for large $\tilde z$ yields integrals with a factor $e^{-2\tilde z^{3/2}/3}$ which can be treated using Laplace's method \cite{bender}. The exponential term turns into the $e^{-8/3\chik}$ suppression of the pair production rates which shows up in all combinations of spin and photon polarisation and reflects the fact that pair production behaves like a tunneling process in the semiclassical limit for small $\chik \ll 1$. Distinguishing the case of parallel spins and anti-parallel spins of the generated pair we find
    \begin{align}
        \mathbb R^{\sigma_p,\sigma_p,\tau_k}_\mathrm{NBW} & \sim 
         \frac{\alpha}{b_k} \chik e^{-\frac{8}{3\chik}} \sqrt{\frac{3}{2}}
              \left[
               \frac{1+\tau_k}{2^5}
               + \frac{13(1+\tau_k)}{3\cdot 2^{11}} \chik
               \right. \nonumber \\ 
               & \left. \qquad 
               + \frac{14677 + 11221 \tau_k }{3^2 \cdot 2^{18}
               } \chik^2
              \right] 
        \,, \label{eqn:PCasysmall1} \\
        \mathbb R^{\sigma_p,-\sigma_p,\tau_k}_\mathrm{NBW} &\sim
           \frac{\alpha}{b_k} \chik e^{-\frac{8}{3\chik}} \sqrt{\frac{3}{2}}
             \left[
                \frac{(1+\sigma_p)(1-\tau_k)}{ 16 } \right. \nonumber\\ & \qquad 
                 + \frac{25-\tau_k + 13 \sigma_p (1-\tau_k)}{ 3\cdot 2^{10}} \chik  \nonumber \\
                 & \left. \qquad 
                 - \frac{ 707 + 5005 \tau_k  - 565 \sigma_p(1 - \tau_k) }{3^2 \cdot 2^{17}}
                 \chik^2
                 \right] \,, \label{eqn:PCasysmall2}
    \end{align}
    as $\chik\to0$.

    In \figref{fig:PP_asy_small} we illustrate the asymptotic limits for small $\chik\ll1$ and scaling of the expressions Eqns.~\eqref{eqn:PCasysmall1}--\eqref{eqn:PCasysmall2}. Here some interesting observations can be made. In all cases, irrespective of the spin-alignment of the pair, the leading order contribution of the NBW rate \emph{explicitly} depends on the incident photon polarisation. For all channels there is an overall exponential suppression factor $e^{-8/3\chik}$ at small $\chik \ll1$, reflecting the tunnelling nature of the NBW pair production process for small $\chik$. It is quite interesting, however, that the exact leading order scaling is very much dependent on the specific channel.
    For parallel spins, the leading order for $\parallel$-photons ($\tau_k=+1$) is $\propto \chik e^{-8/3\chik}$, and for $\perp$-photons it is much smaller $\propto \chik^3 e^{-8/3\chik}$ since the first two terms in \eqref{eqn:PCasysmall1} are proportional to $1+\tau_k$. Moreover, there is no spin-splitting, i.e. the cases $\up\,\up$ and $\down\,\down$ have the same rate. This is in fact true not only for small $\chik$ as can be seen for instance in Fig.~\ref{fig:PPtotal}.

    For anti-parallel spins the leading order of the rate is even more involved. For $\perp$-photons the leading order depends on the spin alignment of the positron. For positrons produced in the favourable $\up$ state, the rate is large, $\propto \chik e^{-8/3\chik}$. However, for positrons produced in the (unfavourable) $\down$ state, the leading order is much smaller at $\propto \chik^2 e^{-8/3\chik}$. This asymptotic result reconfirms the dominance of the $\up\,\down\,\perp$ channel in Fig.~\ref{fig:PP1panel} for $\chik=0.01$. For $\parallel$-photons the leading order for anti-parallel spins is $\propto \chik^2 e^{-8/3\chik}$, independent of the spin alignment of the positron.

    It is know from the literature that in the limit $\chik\ll1$ the pair production rate of $\perp$-photons ($\tau_k=-1$) is twice as large as the rate of $\parallel$-photons \cite{ritus85}. Here we have shown that the former case is dominated by the single spin-polarisation channel $\uparrow\,\downarrow\,\perp$. In contrast, for $\parallel$-photons two equally probable channels contribute. It is also interesting to look at certain ratios of the pair production rates for specific incident photon polarisation. For instance, for $\parallel$-photons, $\tau_k=+1$, the probability to generate the pair with anti-parallel spins is suppressed as $\mathbb R^{\sigma_p,-\sigma_p,+1}_\mathrm{NBW}/ \mathbb R^{\sigma_p,\sigma_p,+1}_\mathrm{NBW} \sim \chik/8$ which is independent of the value of $\sigma_p$. For $\perp$-photons, $\tau_k=-1$, we have to distinguish two cases: $\mathbb R^{\sigma_p,\sigma_p,-1}_\mathrm{NBW}/\mathbb R^{1,-1,-1}_\mathrm{NBW}\sim 3\chik^2/512$ and $\mathbb R^{\sigma_p,\sigma_p,-1}_\mathrm{NBW}/\mathbb R^{-1,1,-1}_\mathrm{NBW}\sim 6/11$.

    The lower plot in Fig.~\ref{fig:PP_asy_small} shows that the asymptotic expressions approximate the NBW rates with a high accuracy only at extremely small values of $\chik\ll1$, and in particular in the interesting range $0.1<\chik<1$ the relative error can be quite large for some channels. (It should be noted that below $\chik<0.1$ the NBW rate is significantly suppressed because of the exponential factor, see Fig.~\ref{fig:PPtotal}.)

    \begin{figure}[!bht]
        \centering
        \includegraphics[width=\columnwidth]{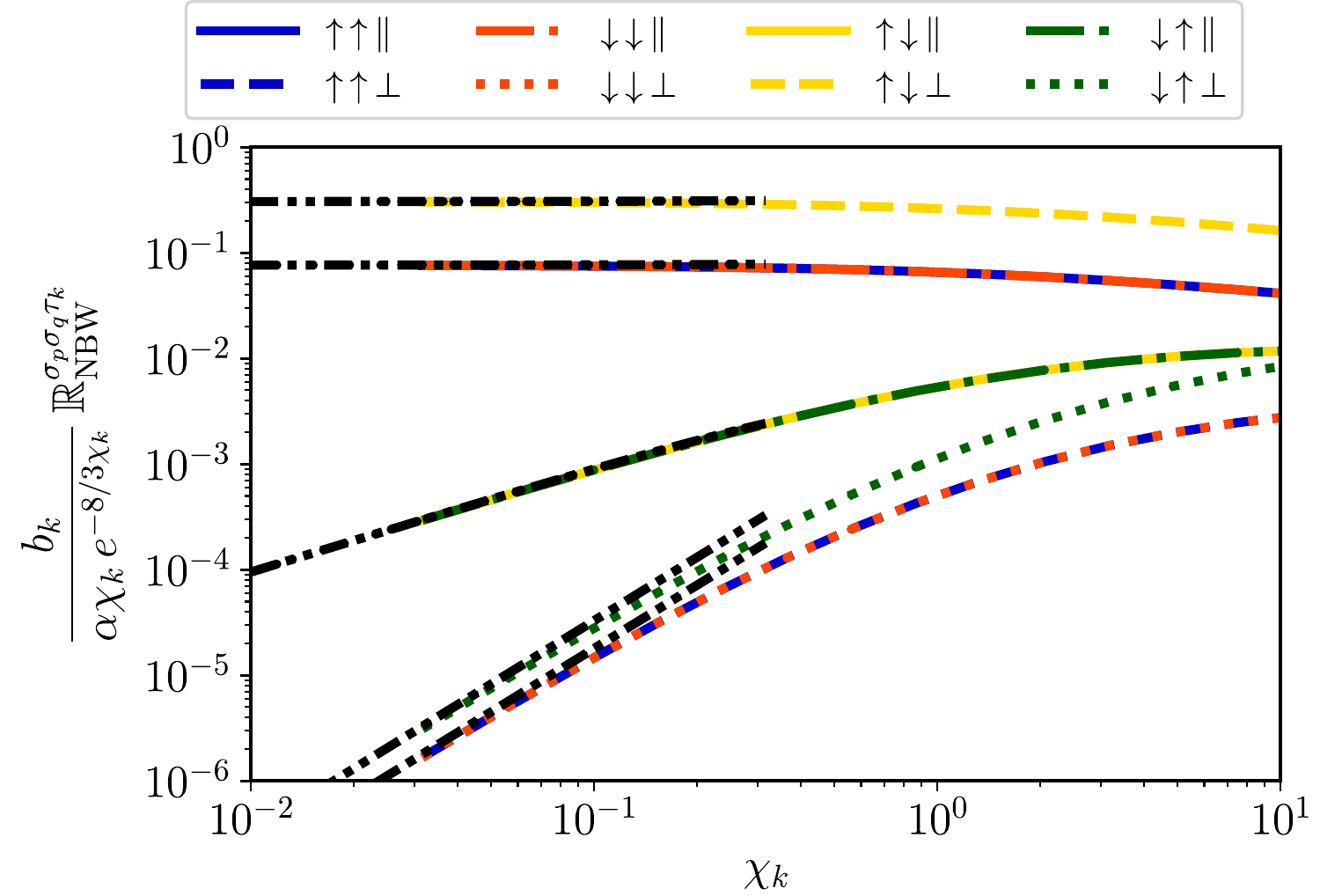}
        \includegraphics[width=\columnwidth]{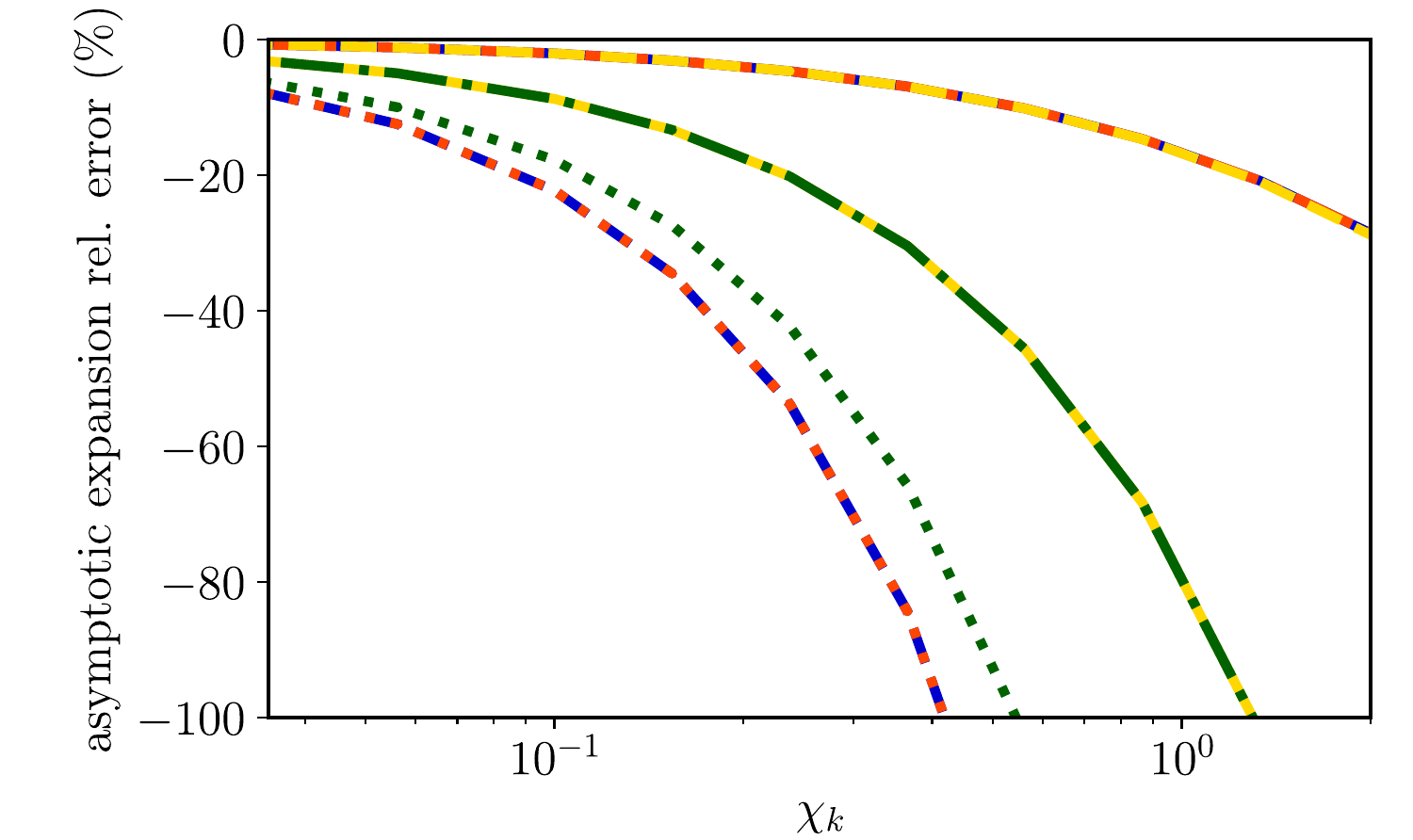}
        \caption{Asymptotics for spin/polarisation resolved NBW pair production rates for small $\chik\ll1$ (black dash-dotted curves) in comparison to the full LCFA rates (top) and relative error of the leading order asymptotic expansion (bottom).}
        \label{fig:PP_asy_small}
    \end{figure}

    \subsubsection{$\chik \gg1$}

    The asymptotic expansion of NBW pair creation for large $\chik$ is calculated in a similar manner as the corresponding NLC expressions.
    The asymptotic expressions behave as
    \begin{align}
    \mathbb R^{\sigma_p,\sigma_p,\tau_k}_\mathrm{NBW} & \sim
        \frac{\alpha \chik^{2/3}}{b_k} 
        \frac{ 3 \cdot 3^{2/3} }{14 \cdot 2^{2/3}}
        \frac{\Gamma(\frac{5}{6})}{\Gamma(\frac{1}{6})}
         \left( 1 + \frac{\tau_k}{2} \right)  \,, \label{eqn:PCasybig1} \\
    \mathbb R^{\sigma_p,-\sigma_p,\tau_k}_\mathrm{NBW} & \sim
    \frac{\alpha \chik^{2/3}}{b_k} 
        \frac{ 3^{2/3}}{ 2\cdot 2^{2/3}}
        \frac{ \Gamma( \frac{5}{6}) }{\Gamma(\frac{1}{6})}
        \left[ \vphantom{\frac{1}{1}}
        \left( 1 - \frac{\tau_k}{2} \right) \right. \nonumber \\
        & \qquad \left.
        + \chik^{-1/3}   \, 
        \sigma_p  ( 1 - \tau_k ) 
        \frac{2^{1/3}}{6\cdot 3^{1/3}}
        \frac{\Gamma^2(\frac{1}{6})}{\Gamma^2(\frac{5}{6})}
         \right] \,,\label{eqn:PCasybig2}
    \end{align}
    as $\chik\to\infty$.

    \begin{figure}[!th]
        \centering
        \includegraphics[width=\columnwidth]{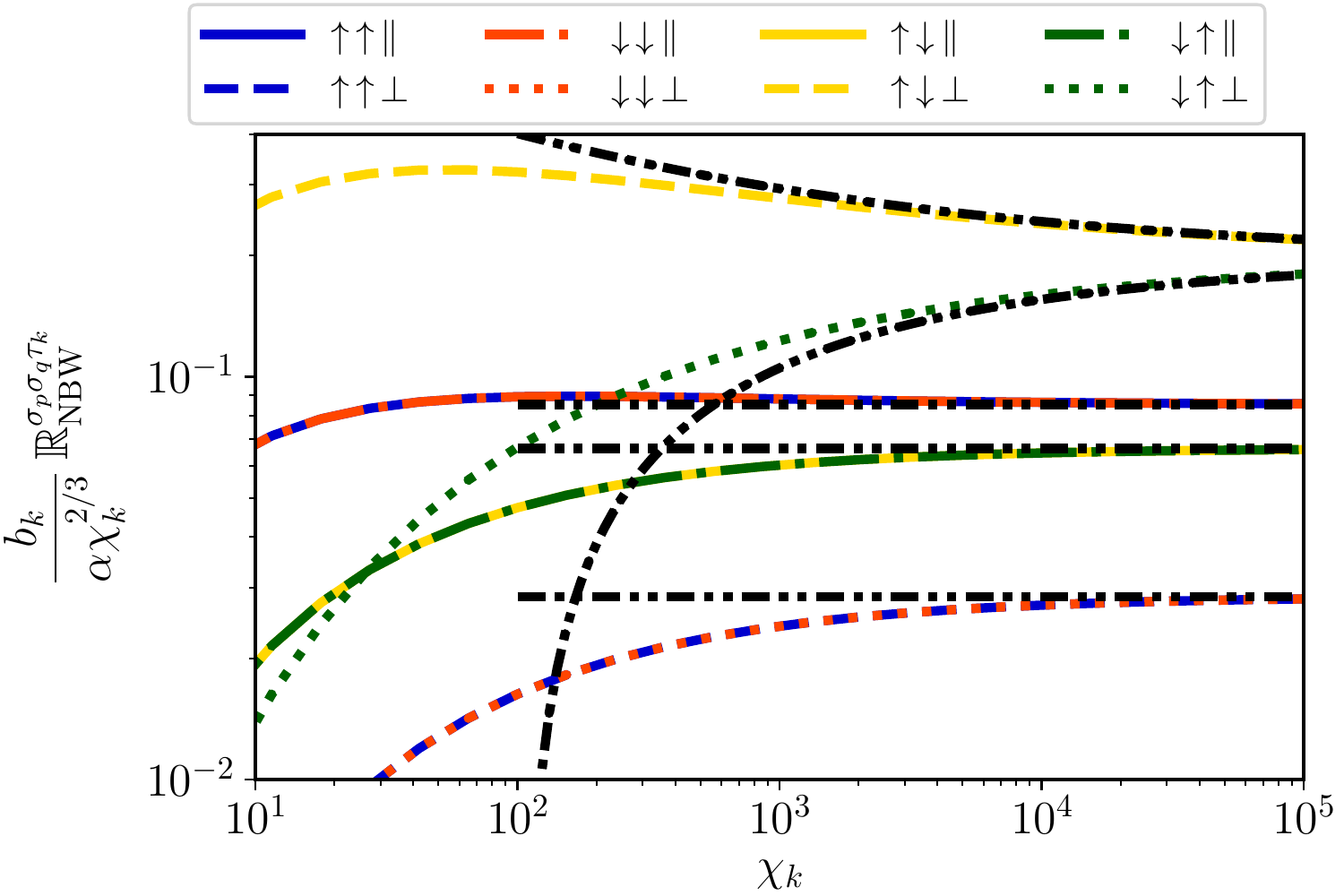}
        \includegraphics[width=\columnwidth]{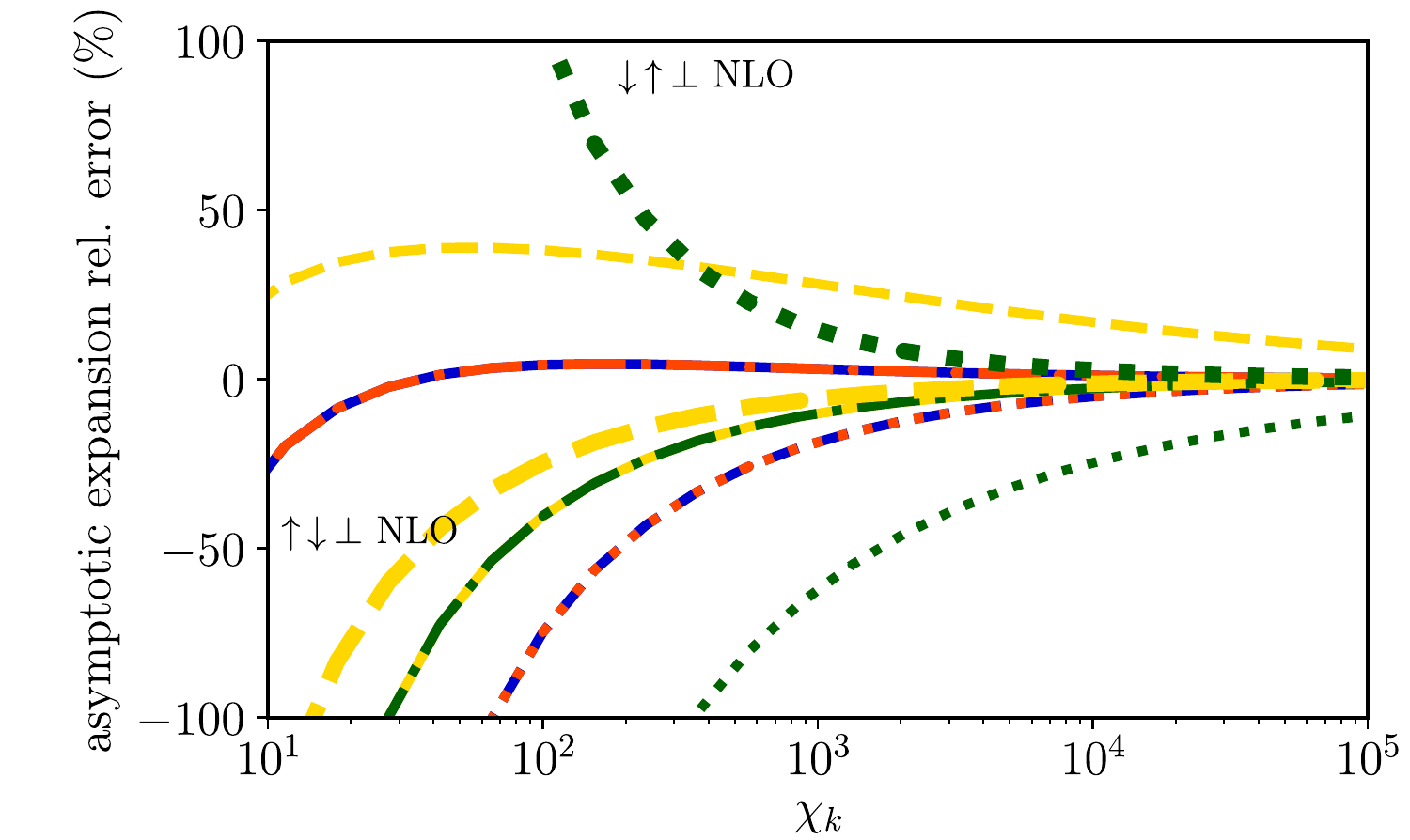}
        \caption{Asymptotics for spin/polarisation resolved pair production rates (black dash-dotted curves) in comparison to the full LCFA for large $\chik\gg1$ (top) and relative error of the asymptotic expansion (bottom) at leading order, except when denoted as NLO.
        }
        \label{fig:PP_asy_large}
    \end{figure}

    Here, the scalings with $\chik$ are in principle the same as for NLC,
    just the numerical factors are different. The main difference is that there is no term at order $\chik^{1/3}$ for the case of parallel spins.
    The asymptotic expressions for large $\chik\gg1$,
    Eqns.~\eqref{eqn:PCasybig1}--\eqref{eqn:PCasybig2} are plotted in Fig.~\ref{fig:PP_asy_large} (top) and the corresponding relative error (bottom).

    The asymptotic plots of the total yield in \figref{fig:PP_asy_small} and \figref{fig:PP_asy_large} also display the behaviour of the ``anomalous'' channel, $\downarrow\,\uparrow\,\perp$. Firstly, it is the only polarisation channel to cross the others, being the equal least probable channel in the $\chik \to 0$ limit since the first two terms in Eq.~\eqref{eqn:PCasysmall2} vanish, but increasing in importance as $\chik$ is increased until the $\chik \to \infty$ limit where it is as probable as the most probable channel. It is remarkable that even by $\chik$ as large as $O(10^5)$, it has not yet reached its asymptotic value. This fact becomes particularly  clear by looking at the relative error of the asymptotic expansions in the bottom panels of Figs.~\ref{fig:PP_asy_small} and \ref{fig:PP_asy_large}. We notice the same behaviour as in the NLC case, that the leading order asymptotic expressions are more accurate already at less extreme asymptotic parameter, whereas the less probable channels require much larger (smaller) values of $\chik$ to reach a given accuracy in the $\chik\to \infty$ ($\chik \to 0$) limits.

    \section{Summary}

    In this paper we have given a comprehensive overview of the rates of two of the most important strong-field QED processes with the polarisation of all particles taken into account. We introduced expressions for fully polarised nonlinear Compton scattering (NLC) and nonlinear Breit-Wheeler pair-creation (NBW) in a general plane-wave background and derived concise formulas for the fully polarised locally constant field approximation (LCFA) of each process. The asymptotic scaling for each process and all of the eight polarisation channels has been derived and presented in succinct expressions, and this scaling has been benchmarked against the full LCFA result. Although some of these results exist in other works in the literature, this is, to the best of our knowledge, the first complete presentation and in-depth analysis of all polarisation channels together. In doing so, we have been able to resolve particle spectra by polarisation channel, and have demonstrated that certain spectral features (such as the appearance of a ``UV shoulder/peak'' at large quantum parameter), are particular to specific polarisation channels. We have also identified ''anomalous`` channels that change in relative importance as the corresponding quantum parameter is increased.

    We note from our results that some polarisation channels do not reach their large-$\chi$ asymptotic scaling until $\chi\gtrsim \mathcal  O(10^{3})$. The Narozhny-Ritus conjecture predicts a breakdown of the QED perturbation expansion in dressed vertices when $\alpha\chi^{2/3}\sim \mathcal O(1)$ \cite{fedotov17,dipiazza19a,ilderton19a,mironov20}, i.e. $\chi \sim \mathcal O(10^{3})$. Furthermore, polarised one-vertex tree-level processes such as in NLC and NBW are necessary in order to correctly factorise higher-order tree-level processes in this perturbation expansion. Therefore it is likely that the resolution of the Narozhny-Ritus conjecture has implications for the relative importance of polarisation channels in NLC and NBW at large $\chi$.

    All our results have been expressed in a polarisation basis that respects the symmetry of the background field. However, depending on how polarisation is measured in experiment, the polarisation of any ``detector'' must be borne in mind. For example, a measurement of high energy photon polarisation has been suggested, which uses the polarisation-dependent probabilities for Bethe-Heitler pair-creation in a Coulomb field \cite{ozaki16,nakamiya17}. Therefore it is the projection of our results onto the natural basis of the Bethe-Heitler polarimeter, which will play a role in any detection. The measurement of the spin-polarisation of high-energy electrons is often performed using M\o{}ller polarimeters \cite{band97,sinclair98}, which, however, are most sensitive to lontigudinal polarisation, or Compton polarimeters \cite{narayan16,barber93} which exploit angular asymmetries in the scattering spectra of \emph{linear} Compton scattering. Some authors also propose to use nonlinear QED processes themselves for polarimetry applications \cite{li19b,wan20}. A review for existing and future electron beam polarimetry can be found in Ref.~\cite{aulenbacher18}.

    Even if the polarisation of the incoming or outgoing particle is not measured, then the  LCFA rates for the eight different polarisation channels we have presented are still relevant for higher-order processes. The correct factorisation of higher-order processes require a consistent polarisation of intermediate particles (propagators) between vertices. In this way, the polarised LCFA rates presented here can be directly employed in numerical simulations of electromagnetic cascades in intense background fields \cite{seipt20}.

 
 	\begin{acknowledgments}
	 B.K. acknowledges support from the EPSRC, Grant No. EP/ S010319/1. 
 	\end{acknowledgments}

    \appendix

    \section{Details of the Calculation of the LCFA for Nonlinear Compton}

    \label{app:nlc}

    We start by giving some important kinematic definitions:
    \begin{align}
    s & \equiv \frac{\kappa . k}{\kappa .p} \,, \\
    g & =   1 + \frac{s^2}{2(1-s)}
    \end{align}

    With help of the auxiliary variable $L$, we can find some useful kinematic relations for the incident electron momentum $p$, outgoing electron momentum $q$ and emitted photon momentum $k$,
    \begin{align}
    p.q &= m^2 + L s \: \kappa. p \,, \\
    q.k &= L \: \kappa.p \,, \\
    p.k &=  L (1-s) \: \kappa.p  \,,
    \end{align}
    where
    \begin{multline}
    L    = \frac{ s}{2 \kappa.p (1-s)  } \left[  m^2  + 
    \frac{X_\varepsilon^2 + X_\beta^2}{s^2} \right] \\
    = x_0 \left[  1 + \left( \frac{\vec p_\perp}{ m } - \vec r_\perp\right)^2 \right] \,,
    \end{multline}
    and we introduced the normalised transverse momentum of the photon, $\vec r_\perp = \vec k_\perp/ms$, and the auxiliary variables $X_\varepsilon = k.\varepsilon-sp.\varepsilon$ and $X_\beta =k.\beta-sp.\beta$. In addition,
    \begin{align} \label{eq:x0}
    x_0 = \frac{s}{2b_p(1-s)} \,,
    \end{align}
    with $b_p = \kappa.p /m^2$.

    \subsection{NLC Traces}

    Here we list the expressions for all 8 Dirac traces for nonlinear Compton scattering, Eqs.~\eqref{eq:NLCtraceUP}--\eqref{eq:NLCtracePC}. They are evaluated using \textsc{FeynCalc} \cite{feyncalc,feyncalc93}. Here we use the short-hand notation $h'$ for $h(\phi')$ and $h$ for $h(\phi)$, and also write $h-h' = \int_{\phi'}^\phi \dot h(\varphi) \ud \varphi = \theta \langle \dot h \rangle $, with $\theta = \phi - \phi'$ being the laser phase difference between the NLC amplitude and its complex conjugate.

    \begin{align}
    \UP_1 &= q.p -m^2     -\frac{m^2 \xi^2  (s-2)^2 }{2 (s-1)} h  h' 
        \nonumber \\ 
        & \qquad 
        +\frac{m \xi  (s-2)^2 }{2 (s-1) s} X_\varepsilon (h+h')
        +\frac{2 X_\varepsilon^2}{s^2} \,, \\
    \UP_2 &=  -q.p - m^2 + 2\frac{k.q}{s}
    +2\frac{(1-s) k.p }{s}
    -\frac{m^2 \xi ^2 s^2 }{2 (s-1)} hh' 
    \nonumber \\ 
        & \qquad 
    + \frac{m \xi  s }{2 (s-1)} X_\varepsilon (h+h')
    -\frac{2}{s^2} X_\varepsilon^2 \,, \\
    \IP_1  &=   i \xi m^2  \, \theta \langle \dot h \rangle \, \frac{s(2-s)}{2(1-s)}\,,\\
    \IP_2  &=  -  i \xi m^2  \, \theta \langle \dot h \rangle \, \frac{s^2}{2(1-s)} \,, \\
    \FP_1 &=  i \xi m^2  \, \theta \langle \dot h \rangle \, \frac{s(2-s)}{2(1-s)}
    \,, \\
    \FP_2 & =    i \xi m^2  \, \theta \langle \dot h \rangle \, \frac{s^2}{2(1-s)} \,, \\
    \PC_1 & = q.p -m^2
    -\frac{m^2 \xi ^2 (s-2)^2 }{2 (s-1)} hh'
        \nonumber \\ 
        & \qquad 
    +\frac{m \xi  (s-2)^2 }{2 (s-1) s} X_\varepsilon( h+h' )
    + \frac{2}{s^2}X_\varepsilon^2 +\frac{X_\beta^2}{s-1} \,,\\
    \PC_2 &= -3 q.p + m^2
    + \frac{2 k.q}{s}
    + \frac{2 (1-s) k.p }{s}
        +\frac{m^2 \xi ^2 s^2 }{2 (s-1)} h h'
        \nonumber \\ 
        & \qquad 
        -\frac{m \xi  s }{2 (s-1)} X_\varepsilon ( h + h' ) 
        -\frac{2}{s^2}X_\varepsilon^2 -\frac{X_\beta^2}{s-1}
    \end{align}
    By using the kinematic relations from above some of the expressions can be simplified, e.g.~$q.p-m^2 = Ls \kappa.p$. With these replacements it is straightforward to see that all traces depend on the transverse photon momentum only quadratically at most.
	Here we used that the light-front Levi-Cevita tensor $\epsilon^{+-xy} = -2$, i.e.~that Levi-Civita terms occurring in traces with exactly one $\gamma^5$ matrix can be simplified as $\epsilon^{p\beta\epsilon\kappa}=p.\kappa$.

\subsection{Gaussian Transverse Momentum Integrals}
\label{sect:transverse:integral}

    We find that the transverse momentum integrals over $\vec r_\perp$ are all Gaussian for all 8 Compton traces. This fact has been customarily exploited in calculations of spin-averaged nonlinear Compton scattering, to analytically perform the transverse momentum integrals. Here, the relevant integrals for polarised NLC read
    \begin{align}
        \mathscr G_0 
        		& = \int d^2 \vec r_\perp \: e^{i\theta \frac{k. \langle\pi_p\rangle}{\kappa.q}} 
        		  = \frac{\pi}{-i\theta x_0} \: e^{i\theta x_0 \mu}    \,, \\
        \mathscr G_{1,\varepsilon} 
                & = \int d^2 \vec r_\perp \: X_\varepsilon \:e^{i\theta \frac{k. \langle\pi_p\rangle}{\kappa.q}}
        		  = ms\xi \langle h \rangle  \mathscr G_0 \,, \\
        \mathscr G_{2,\varepsilon} 
                & = \int d^2 \vec r_\perp \: X_\varepsilon^2 \:e^{i\theta \frac{k. \langle\pi_p\rangle}{\kappa.q}}
        		  =  m^2 s^2 \left[ \xi^2 \langle h\rangle^2 + \frac{1}{-2i\theta x_0} \right] \mathscr G_0 \,, \\
        \mathscr G_{1,\beta} 
                & = \int d^2 \vec r_\perp \: X_\beta \:e^{i\theta \frac{k. \langle\pi_p\rangle}{\kappa.q}}
                  = 0 \,, \\
        \mathscr G_{2,\beta} 
                & = \int d^2 \vec r_\perp \: X_\beta^2 \:e^{i\theta \frac{k. \langle\pi_p\rangle}{\kappa.q}}
                  =   \frac{m^2 s^2}{-2i\theta x_0} \mathscr G_0 \,,
    \end{align}
    with $x_0$ defined in Eq.~\eqref{eq:x0}. With these, we find (we omit the leading factor $\mathscr G_0$ here which has to be multiplied to all traces)
    \begin{align} \label{eq:UP1-theta21} \allowdisplaybreaks
        \UP_1 &
                \to (g-1) m^2 + i \frac{gm^2}{\theta x_0}  
                \nonumber \\ & \qquad
                + m^2 \xi^2 (g+1) (h- \langle h\rangle) (h' - \langle h \rangle ) \,, \\
        \UP_2 &
                \to (g-1) m^2 + i \frac{gm^2}{\theta x_0}  
                \nonumber \\ & \qquad
                + m^2 \xi^2 (g-1) (h- \langle h\rangle) (h' - \langle h \rangle ) \,, \displaybreak \\
        \IP_1  &
                \to   i \xi m^2  \, \theta \langle \dot h \rangle \, (g-1+s) \,,\\
        \IP_2  &
                \to  -  i \xi m^2  \, \theta \langle \dot h \rangle \, (g-1) \,,  \\
        \FP_1 &
                \to  i \xi m^2  \, \theta \langle \dot h \rangle \, (g-1+s) \,, \\
        \FP_2 & 
                \to  i \xi m^2  \, \theta \langle \dot h \rangle \, (g-1) \,, \\
        \PC_1 &
                \to (g-1) m^2 + i \frac{m^2}{\theta x_0}  
                \nonumber \\ & \qquad
                + m^2 \xi^2 (g+1) (h- \langle h\rangle) (h' - \langle h \rangle ) \,, \\
        \PC_2 &
                \to - (g-1) m^2 + i \frac{m^2}{\theta x_0}  
                \nonumber \\ & \qquad
                - m^2 \xi^2 (g-1) (h- \langle h\rangle) (h' - \langle h \rangle ) \,.
    \end{align}

    \subsection{Short Coherence Interval Approximation and $\theta$-integrals}

    With the transverse momentum integrals done, the next step towards the LCFA is
    to expand the integrand of the $\theta$-integral to lowest non-trivial order in the short coherence interval $\theta\ll1$. This allows us to perform the $\theta$-integrals analytically. (Note that one can alternatively perform the $\theta$-integral first, and not perform the $\vec r_\perp$ integrals, which leads to an angularly resolved LCFA, see for instance Ref.~\cite{blackburn20}.) For the Kibble mass in the exponent that means $\mu \to \mu_0 =  1 + \xi^2\dot h^2\theta^2/12 $ \cite{king19a}. Furthermore, in the pre-exponential terms we use $\theta \langle \dot h \rangle \to \theta \dot h = \xi^2 \dot h^2(\varphi)$ and
    \begin{align}
	(h' -  \langle h \rangle  )(h-  \langle h \rangle ) &\simeq -\frac{\theta^2}{4} \dot h^2 \,.
    \end{align}
    Inserting the small-$\theta$ approximated prefactor 
     $\mathscr G_0 \simeq 2\pi b_p \, \frac{1-s}{s} \frac{ e^{i\theta x_0 \mu_0} }{-i\theta  }$ we obtain
    \begin{widetext}
    \begin{align} \label{eq:UP1-theta22} \allowdisplaybreaks
    \int \! \ud \vec r_\perp \:e^{i\theta \frac{k. \langle\pi_p\rangle}{\kappa.q}} \UP_1 &  \simeq
    2\pi b_p m^2  \, \frac{1-s}{s}  e^{i\theta x_0 \mu_0}
    \left[
    -\frac{g }{\theta ^2 x_0 } + \frac{i (g-1)}{\theta } - i \theta  (g+1)\frac{ \dot h^2  \xi ^2 }{4}     \right] \,, \\
    \int \! \ud \vec r_\perp \:e^{i\theta \frac{k. \langle\pi_p\rangle}{\kappa.q}} \UP_2 & \simeq
    2\pi b_p m^2  \, \frac{1-s}{s}  e^{i\theta x_0 \mu_0}
    \left[
    -\frac{g }{\theta ^2 x_0 } + \frac{i (g-1)}{\theta } - i \theta  (g-1)\frac{ \dot h^2  \xi ^2 }{4}     \right] \,,  \\
    \int \! \ud^2 \vec r_\perp \:e^{i\theta \frac{k. \langle\pi_p\rangle}{\kappa.q}} \IP_1  
     & \simeq  - 2\pi m^2b_p \frac{1-s}{s} \: e^{i\theta x_0 \mu_0} \: \xi \dot h   \: (g-1+s)  \,,\\%
    \int \! \ud^2 \vec r_\perp \:e^{i\theta \frac{k. \langle\pi_p\rangle}{\kappa.q}} \IP_2 
    &  \simeq   2\pi m^2b_p \frac{1-s}{s} \: e^{i\theta x_0 \mu_0}  \:  \xi   \dot h   \: (g-1)  \,,  \\
    \int \! \ud^2 \vec r_\perp \:e^{i\theta \frac{k. \langle\pi_p\rangle}{\kappa.q}} \FP_1  
     &  \simeq   - 2\pi m^2b_p \frac{1-s}{s} \: e^{i\theta x_0 \mu_0}  \: \xi \dot h  \: (g-1+s) \,, \\
    \int \! \ud^2 \vec r_\perp \:e^{i\theta \frac{k. \langle\pi_p\rangle}{\kappa.q}} \FP_2 
     & \simeq  -2\pi m^2b_p \frac{1-s}{s} \: e^{i\theta x_0 \mu_0}     \: \xi   \dot h  \: (g-1) \,, \\
     \int \! \ud \vec r_\perp \:e^{i\theta \frac{k. \langle\pi_p\rangle}{\kappa.q}} \PC_1 &  \simeq
    2\pi b_p m^2  \, \frac{1-s}{s}  e^{i\theta x_0 \mu_0}
    \left[
    -\frac{1 }{\theta ^2 x_0 } + \frac{i (g-1)}{\theta } - i \theta  (g+1)\frac{ \dot h^2  \xi ^2 }{4}     \right] \,, \\
    \int \! \ud \vec r_\perp \:e^{i\theta \frac{k. \langle\pi_p\rangle}{\kappa.q}} \PC_2 & \simeq
    2\pi b_p m^2  \, \frac{1-s}{s}  e^{i\theta x_0 \mu_0}
    \left[
    -\frac{1 }{\theta ^2 x_0 } - \frac{i (g-1)}{\theta } + i \theta  (g-1)\frac{ \dot h^2  \xi ^2 }{4}     \right] \,.
    \end{align}
    \end{widetext}

    Next we perform the integrals over the phase variable $\theta$ yielding Airy functions
    \begin{align} \label{eq:airy1}
    \int \! \ud \theta  \: i\theta \: e^{ix_0 \theta + i \frac{y}{3} \theta^3 } &= 2\pi \frac{\Ai'(z)}{\sqrt[3]{y}^2} \,,\\ 
    \int \! \ud \theta  \:  e^{ix_0 \theta + i \frac{y}{3} \theta^3 } &= 2\pi \frac{\Ai(z)}{\sqrt[3]{y}} \,,\\ 
    \int \! \ud \theta  \: \frac{1}{-i\theta} \:  e^{ix_0 \theta + i \frac{y}{3} \theta^3 } &= 2\pi \Ai_1(z) \,,\\ 
    \int \! \ud \theta \: \frac{1}{\theta^2} \:  e^{ix_0 \theta + i \frac{y}{3} \theta^3 } & 
    = 2\pi  x_0 \left[ \Ai_1(z) +  \frac{\Ai'(z)}{z} \right] \,,
    \label{eq:airy4}
    \end{align}
    where $\Ai_1(z)=\int_z^\infty \! \ud x  \: \Ai(x)$ and $\Ai'(z)  = \ud \Ai (z)/\ud z$.
    
    Here we have rewritten the exponential $e^{i\theta x_0 \mu_0 } = e^{i x_0 \theta + i\frac{y}{3}\theta^3}$ with the definitions
    \begin{align}
	y    &= \frac{x_0 \xi^2 \dot h^2}{4} \,, 
			\qquad  
	z    = \frac{x_0}{\sqrt[3]{y}}   \,.
    \end{align}
	In addition we use that $\sqrt[3]{y} = \sqrt{z}\xi |\dot h| / 2 $ and thus
	$\xi \dot h / \sqrt[3]{y}
	= 2 \dot h / ( \sqrt{ z} |\dot h|) 
	=  2 \, \shd / \sqrt{ z}  $.

    The first and second results follow by the integral definition of the Airy function \cite{soares10}. The third result can be derived in the following way:
    \begin{multline}
    \int_{-\infty}^{\infty}\frac{\ud \theta}{\theta} e^{i(r\theta + c_{3}\theta^{3})} = \lim_{\eps\to0} \int_{-\infty}^{\infty}\frac{\ud \theta}{\theta+i\eps} e^{i(r\theta + c_{3}\theta^{3})}\\
    = \lim_{\eps\to0} -i\int_{0}^{\infty}\ud v\int_{-\infty}^{\infty}\frac{\ud \theta}{\theta+i\eps} e^{i((r+v)\theta + c_{3}\theta^{3})-\eps\,v\theta} \\
    = - 2\pi i \Ai_{1}\left[\frac{r}{(3c_{3})^{1/3}}\right],
    \end{multline}
    and the final result was derived in the appendix of Ref.~\cite{king20a}. (It turns out the final result is equivalent to integrating once by parts, ignoring the contribution from the pole in the evaluated term, and then using the standard Sokhotsky-Weierstrass method to deal with the pole of the resulting $1/\theta$ integration.)

    Here is the collection of all 8 NLC traces after the $\theta$-integrals have been performed:
    \begin{widetext}
    \begin{align}
        \int \! \ud \theta \int \! \ud^2 \vec r_\perp \:e^{i\theta \frac{k. \langle\pi_p\rangle}{\kappa.q}}  
        \UP_1  
            & \simeq  - 4\pi^2 m^2b_p \frac{1-s}{s}  \left[ \Ai_1(z)+  \frac{2g+1}{z} \Ai'(z)		\right]	\,,\\
        \int \! \ud \theta \int \! \ud^2 \vec r_\perp \:e^{i\theta \frac{k. \langle\pi_p\rangle}{\kappa.q}} 
        \UP_2 
            & \simeq - 4\pi^2 m^2b_p \frac{1-s}{s} \left[   \Ai_1(z) + \frac{2g-1}{z} \Ai'(z) \right]  \,, \\
        \int \! \ud \theta\int \! \ud^2 \vec r_\perp \:e^{i\theta \frac{k. \langle\pi_p\rangle}{\kappa.q}}  
        \IP_1  
            & \simeq   - 4\pi^2 m^2b_p \frac{1-s}{s}  \: (g-1+s) \: \frac{2\Ai(z)}{\sqrt{z}}  \: \shd\,,\\
        \int \! \ud \theta\int \! \ud^2 \vec r_\perp \:e^{i\theta \frac{k. \langle\pi_p\rangle}{\kappa.q}} 
        \IP _2 
            & \simeq  -  4\pi^2 m^2b_p \frac{1-s}{s} \:   (1-g) \: \frac{2\Ai(z)}{\sqrt{z}} \: \shd\,, \\
        \int \! \ud \theta\int \! \ud^2 \vec r_\perp \:e^{i\theta \frac{k. \langle\pi_p\rangle}{\kappa.q}}  
        \FP_1  
            &  \simeq  - 4\pi^2 m^2b_p \frac{1-s}{s}   \: (g-1+s) \: \frac{2\Ai(z)}{\sqrt{z}} \: \shd\,, \\
        \int \! \ud \theta\int \! \ud^2 \vec r_\perp  \:e^{i\theta \frac{k. \langle\pi_p\rangle}{\kappa.q}}  
        \FP_2 
            &  \simeq - 4\pi^2 m^2b_p \frac{1-s}{s}       \: (g-1) \: \frac{2\Ai(z)}{\sqrt{z}}\: \shd \,, \\
        \int \! \ud \theta \int \! \ud^2 \vec r_\perp \:e^{i\theta \frac{k. \langle\pi_p\rangle}{\kappa.q}}   \
        \PC_1 
            & \simeq - 4\pi^2 m^2b_p \frac{1-s}{s} \left[  ( 2 -g )  \: \Ai_1(z)  +  \frac{g+2}{z}  \: \Ai'(z)    \right] \,,\\
        \int \! \ud \theta \int \! \ud^2 \vec r_\perp \:e^{i\theta \frac{k. \langle\pi_p\rangle}{\kappa.q}}   
        \PC_2
        	& \simeq -4\pi^2 m^2b_p \frac{1-s}{s} \left[  g \, \Ai_1(z) -  \frac{ g-2}{z}  \: \Ai'(z)  \right] \,.
    \end{align}
    By combining these results according to Eqn.~\eqref{eq:Pj_NLC}, and by defining the differential
    probability rate per laser phase as $\ud \mathbb R/\ud s = \ud \mathbb P/\ud s\ud \varphi$ we find
    \begin{align} \label{eq:RNLC1}
    \frac{\ud \mathbb R_\mathrm{NLC,1}}{\ud s}( \sigma_p ,  \sigma_{q} )		
    	 &= 
    	 -  \frac{\alpha}{4b_p} 
    		 \left[ 
    		 (1 +  \sigma_p \sigma_{q} (2-g) ) \Ai_1(z) 
    		 + 2( \sigma_p + \sigma_{q})(g-1+s) \frac{\Ai(z)}{\sqrt{z}} \: \shd \right. \nonumber  \\
    		& \qquad \qquad \quad 
    		    + \left. \left( 2g+1 +  \sigma_p \sigma_{q}  (g+2) \right) \frac{\Ai'(z)}{z}
    		 \right] \,, \\
    		\label{eq:RNLC2}
    \frac{\ud \mathbb R_\mathrm{NLC,2}}{\ud s}	( \sigma_p , \sigma_{q} )		
    	& =	
    		 -  \frac{\alpha}{4b_p} 
    	\left[ 
    	(1 +  \sigma_p \sigma_{q}  g ) \Ai_1(z) 
    	+ 2(\sigma_{q} - \sigma_{p}) (g-1) \frac{\Ai(z)}{\sqrt{z}} \: \shd \right. \nonumber \\
    	& \qquad \qquad \quad 
    	+ \left. \left( 2g-1 - \sigma_p \sigma_{q} (g-2) \right) \frac{\Ai'(z)}{z}
    	\right] \,.
    \end{align}
    for a photon to be emitted in polarisation state $\Lambda_1$ or $\Lambda_2$.
    \end{widetext}

    \section{Details of the Calculation of the LCFA for Pair Production}
    \label{app:pp}

    For pair production, the incoming channel is characterised by the scalar product $\kappa.k$, with $k$ the photon four-momentum. The light-front momentum exchange is defined here as $s = p.\kappa/k.\kappa$, where $p$ refers to the \emph{positron} momentum. Hence, for the electron momentum $q$ we have $q.\kappa = (1-s) q.\kappa $. Moreover, we define $\tilde g = 1 - \frac{1}{2s(1-s)}$.

    By introducing the auxiliary variable $\tilde L$, it is possible to express
    \begin{align}
    p.q &= \tilde L k.\kappa - m^2 \,, \\
    q.k &= \tilde L s k.\kappa \,,\\
    p.k &= \tilde L (1-s) k.\kappa \,,
    \end{align}
    where
    \begin{align}
    \tilde L & = 
     \tilde x_0\left[  1 + s^2 \left(  \vec r_\perp - \frac{\vec k_\perp}{m} \right)^2 \right]
    = \frac{m^2 + Y_\varepsilon^2 + Y_\beta^2}{2s(1-s) k.\kappa }
    \end{align}
    with
    \begin{align} \label{eq:tildex0}
      \tilde x_0 = \frac{1}{2b_k s(1-s)} \,
    \end{align}
    and
    $Y_\varepsilon = p.\varepsilon -s k.\varepsilon $ and $Y_\beta = p.\beta -s k.\beta $, and the normalised transverse positron momentum $\vec r_\perp = \vec p_\perp / ms$. $b_k = k.\kappa/m^2$ is related to the squared centre-of-mass energy of the incident photons and can be related to the kinematic pair production threshold of linear Breit-Wheeler via $\tilde L b_k \geq 2$, or $\tilde L \geq 2/b_k$.

    The dynamic phase of the pair production matrix element reads (without and with the floating average)
    \begin{align}
    \frac{-k.\pi_{-p} }{\kappa.q}&= 
     \frac{1}{2 k.\kappa s(1-s)} \left[ m^2 + (Y_\varepsilon - m \xi h )^2 + Y_\beta^2  \right] \,, \\
     \frac{-k.\langle \pi_{-p} \rangle }{\kappa.q} 
    & = \frac{1}{2 k.\kappa s(1-s)} \left[ m^2 \mu + (Y_\varepsilon - m \xi \langle h \rangle  )^2 + Y_\beta^2  \right] \nonumber \\ 
    &= 
     \tilde x_0 \left[ \mu 
    + s^2 ( \vec r_\perp + \langle \vec a_\perp \rangle /s -  \vec u_\perp )^2 \right]\,, 
   \end{align}
    with the Kibble mass $\mu$, Eq.~\eqref{eq:Kibblemass}.

    \subsection{NBW Traces}

    The NBW, Eqs.~\eqref{eq:NBWtraceUP}--\eqref{eq:NBWtracePC}, traces are calculated in an analogous way to the NLC traces,
    \begin{align}
    \UP_1 & = q.p + m^2
    -\frac{m^2 \xi ^2 (1-2 s)^2 }{2 (s-1) s} hh'
     \nonumber \\ & \qquad 
    + \frac{m \xi  (1-2 s)^2 }{2 (s-1) s} (h+h') Y_\varepsilon 
    -2 Y_\varepsilon^2  \,, \\
    \UP_2 & = -q.p + m^2 
        +2 (1-s) k.p
        +2 s k.q
        -\frac{m^2 \xi ^2 }{2 (s-1) s} hh'
     \nonumber \\ & \qquad 
    + \frac{m \xi }{2 (s-1) s}(h+h')Y_\varepsilon 
    +2 Y_\varepsilon^2 \,,\\
        \PP_1 
    	&= i  m^2 \xi  \:  \theta \langle \dot h \rangle \: 
                \frac{ 2 s-1 }{2 s (1-s) } \,, \\
    \PP_2&=  - i m^2 \xi \: \theta \langle \dot h \rangle \:  
      	        \frac{ 1 }{2 s (1-s)  } \,, \\
    \EP_1 &= i m^2 \xi \: \theta \langle \dot h \rangle \:  \frac{ 2 s-1}{2 s(1-s) } \,, \\
    \EP_2 &= i m^2 \xi \: \theta \langle \dot h \rangle \:  \frac{ 1 }{2 s (1-s) } \,,\\
    \PC_1 &= q.p + m^2 
        -\frac{m^2 \xi ^2 (1-2 s)^2 }{2 (s-1) s} hh'
     \nonumber \\ & \qquad 
        +\frac{m \xi  (1-2 s)^2 }{2 (s-1) s} (h+h') Y_\varepsilon
        -2 Y_\varepsilon^2 +\frac{Y_\beta ^2}{(s-1) s} \,,\\
    \PC_2 &= -3 q.p -m^2
             +2 (1-s) k.p
             +2 s k.q
        +\frac{m^2 \xi ^2 }{2 (s-1) s} hh'
     \nonumber \\ & \qquad 
    - \frac{m \xi }{2 (s-1) s}(h+h') Y_\varepsilon
    +2 Y_\varepsilon^2 -\frac{Y_\beta^2}{(s-1) s}
    \,.
    \end{align}
    By employing the kinematic relations from above it is straightforward to see that all transverse momentum integrals over the eight traces are Gaussian.

    \subsection{Gaussian Transverse Momentum Integrals for NBW}
    \label{sect:transverse:integral:BWPP}

For NBW pair production, all transverse momentum integrals over $\ud^2 \vec r_\perp$ are Gaussian as well. However, the expression of the dynamic phase
is slightly different, and so are the results:
\begin{align}
\tilde {\mathscr G}_0  & = \int d^2 \vec r_\perp \: e^{i\theta \frac{-k. \langle\pi_{-p}\rangle}{\kappa.q}} 
= e^{i\theta \tilde x_0 \mu } \frac{\pi}{-i\theta \tilde x_0 s^2}   \,, \\
\tilde {\mathscr G}_{1,\varepsilon} 
&= \int d^2 \vec r_\perp \: Y_\varepsilon \:e^{i\theta \frac{- k. \langle\pi_{-p}\rangle}{\kappa.q}}
= m\xi \langle h \rangle  \tilde{\mathscr G}_0 \,, \\
\tilde {\mathscr G}_{2,\varepsilon} &= \int d^2 \vec r_\perp \: Y_\varepsilon^2 \:e^{i\theta \frac{-k. \langle\pi_{-p}\rangle}{\kappa.q}}
=  \left[  m^2 \xi^2 \langle h\rangle^2 + \frac{m^2 }{-2i \tilde \theta x_0} \right]  \tilde{\mathscr G}_0 \,, \\
\tilde{ \mathscr G}_{1,\beta} 
&= \int d^2 \vec r_\perp \: Y_\beta \:e^{i\theta \frac{k. \langle\pi\rangle}{\kappa.q}} = 0 \,, \\
\tilde{\mathscr G}_{2,\beta} &= \int d^2 \vec r_\perp \: Y_\beta^2 \:e^{i\theta \frac{k. \langle\pi\rangle}{\kappa.q}}
=   \frac{m^2}{-2i\theta \tilde x_0} \tilde{ \mathscr G}_0 \,,
\end{align}
with $\tilde x_0$ defined in Eq.~\eqref{eq:tildex0}.
Employing those Gaussian integrals, the 8 NBW traces turn to
the following expressions, omitting again the leading factor 
$\tilde{ \mathscr G}_0$:
\begin{align}
\UP_1 & \to   ( 1 - \tilde g ) m^2 
    -\frac{i \tilde g m^2}{\theta  \tilde x_0} 
    \nonumber \\ & \qquad 
            - (1+ \tilde g )  m^2 \xi ^2 
                (h - \langle h\rangle) (h' -\langle h\rangle )
             \,, \\
\UP_2 & \to     (1- \tilde g) m^2
 -\frac{i \tilde g m^2}{\theta  \tilde x_0 }
    \nonumber \\ & \qquad 
            + ( 1 - \tilde g ) m^2 \xi ^2  
            (h - \langle h\rangle) (h' -\langle h\rangle )
            \,, \displaybreak \\
\PP_1 & \to - i  m^2 \xi  \:  \theta \langle \dot h \rangle \: 
	    (\tilde g-1+s^{-1}) \,, \\
\PP_2&\to  i m^2 \xi \: \theta \langle \dot h \rangle \:   (\tilde g-1) \,, \\
\EP_1 &\to - i m^2 \xi \: \theta \langle \dot h \rangle \: (\tilde g-1+s^{-1}) \,, \\
\EP_2 &\to - i m^2 \xi \: \theta \langle \dot h \rangle \:  (\tilde g-1 ) \,,\\
\PC_1 &\to ( 1 - \tilde g ) m^2
        -\frac{i m^2}{\theta  \tilde x_0 }
    \nonumber \\ & \qquad 
        -( 1 + \tilde g ) m^2 \xi ^2
            (h - \langle h\rangle) (h' -\langle h\rangle )
\,,\\
\PC_2 & \to -(1- \tilde g) m^2
            -\frac{i m^2}{\theta \tilde x_0} \,.
        \nonumber \\ & \qquad 
            -(1- \tilde g) m^2 \xi ^2        
            (h - \langle h\rangle) (h' -\langle h\rangle )
\end{align}

\subsection{Short Coherence Interval Approximation and $\theta$-Integrals}

The next step towards the LCFA for NBW is approximating the integrand for short coherence interval $\theta \ll 1$. This is exactly the same as for NLC.
The only notable difference is that we have to insert here the small-$\theta$ approximation of 
$\tilde {\mathscr G}_0 \simeq e^{i\theta \tilde x_0 \mu_0 } \frac{\pi}{-i\theta \tilde x_0 s^2} = 2\pi b_k \frac{1-s}{s} \frac{e^{i\theta \tilde x_0 \mu_0 }}{-i\theta} $:
\begin{widetext}
\begin{align}
\int \!\ud^2 \vec r_\perp \: e^{i\theta \frac{-k. \langle\pi_{-p}\rangle}{\kappa.q}}  \: \UP_1 
    & \simeq 2\pi b_k m^2  \frac{1-s}{s} \, e^{i\theta \tilde x_0 \mu_0 } 
        \left[ i \theta \frac{  \xi ^2 \dot h^2}{4} (\tilde g+1)  
        +\frac{i (1-\tilde g) }{\theta }
        +\frac{\tilde g }{\theta ^2 \tilde x_0 } \right] \,, \\
\int \!\ud^2 \vec r_\perp \: e^{i\theta \frac{-k. \langle\pi_{-p}\rangle}{\kappa.q}}  \: \UP_2 
    & \simeq 2\pi b_k m^2  \frac{1-s}{s} \, e^{i\theta \tilde x_0 \mu_0 } 
        \left[ i \theta \frac{\xi ^2 \dot h^2 }{4}  (\tilde g-1)      
               + \frac{i (1-\tilde g)}{\theta }+\frac{\tilde g}{\theta ^2 \tilde x_0} \right] \,,\\
\int \!\ud^2 \vec r_\perp \: e^{i\theta \frac{-k. \langle\pi_{-p}\rangle}{\kappa.q}} \PP_1 
& \simeq 2\pi  m^2 b_k \frac{1-s}{s} \: e^{i\theta \tilde x_0 \mu_0}\:  \xi  \dot h         \:  \left(\tilde g - 1 + \frac{1}{s} \right ) \,, \\
\int \!\ud^2 \vec r_\perp \: e^{i\theta \frac{-k. \langle\pi_{-p}\rangle}{\kappa.q}} \PP_2 
& \simeq  2\pi m^2 b_k  \frac{1-s}{s} \: e^{i\theta \tilde x_0 \mu_0} \: \xi  \dot h  \:  (1 - \tilde g )   \,,\\
\int \!\ud^2 \vec r_\perp \: e^{i\theta \frac{-k. \langle\pi_{-p}\rangle}{\kappa.q}} \EP_1 
& \simeq 2\pi  m^2 b_k \frac{1-s}{s} \: e^{i\theta \tilde x_0 \mu_0}\:  \xi  \dot h         \:  \left(\tilde g - 1 + \frac{1}{s} \right ) \,, \\
\int \!\ud^2 \vec r_\perp \: e^{i\theta \frac{-k. \langle\pi_{-p}\rangle}{\kappa.q}} \EP_2
 &  \simeq 2\pi m^2 b_k  \frac{1-s}{s} \: e^{i\theta \tilde x_0 \mu_0} \: \xi  \dot h  \:  ( \tilde g - 1 )   \,,\\
\int \!\ud^2 \vec r_\perp \: e^{i\theta \frac{-k. \langle\pi_{-p}\rangle}{\kappa.q}}  \: \PC_1 
    & \simeq 2\pi b_k m^2  \frac{1-s}{s} \, e^{i\theta \tilde x_0 \mu_0 } 
        \left[   i\theta \frac{\xi^2 \dot h^2 }{4} ( 1 + \tilde g  ) 
                    + \frac{i (1-\tilde g)}{\theta }
                +\frac{1}{\theta ^2 \tilde x_0 }
                \right] \,,\\
\int \!\ud^2 \vec r_\perp \: e^{i\theta \frac{-k. \langle\pi_{-p}\rangle}{\kappa.q}}  \: \PC_2 
    & \simeq 2\pi b_k m^2  \frac{1-s}{s} \, e^{i\theta \tilde x_0 \mu_0 } 
        \left[
        i\theta \frac{\xi ^2 \dot h^2}{4} (1 - \tilde g)  
        -\frac{i (1-\tilde g)}{\theta }
        +\frac{1}{\theta^2 \tilde x_0}
        \right] \,.
\end{align}

	Next we have to perform the integrals over $\theta$ which will yield the Airy functions. 

	The results of the $\theta$-integration are 
	formally the same as for Compton, Eqns.~\eqref{eq:airy1}--\eqref{eq:airy4},
	but with the replacements $x_0\to \tilde x_0$, $y\to \tilde y$ and $z\to \tilde z$, where
	\begin{align}
	\tilde y    &= \frac{\tilde x_0 \xi^2 \dot h^2}{4}  \,,\qquad
	\tilde z    = \frac{\tilde x_0}{\sqrt[3]{\tilde y}} = \left( \frac{1}{\chi_k |\dot h| s(1-s)} \right)^{2/3} \,.
	\end{align}
	
    With these results we obtain for the 8 NBW pair production traces:
	\begin{align}
	\int \! \ud \theta \int \!\ud^2 \vec r_\perp \: e^{i\theta \frac{-k. \langle\pi_{-p}\rangle}{\kappa.q}} \UP_1 
	&  \simeq  4\pi^2 m^2 b_k \frac{1-s}{s} 
	\left[  \Ai_1(\tilde z) +   \frac{2 \tilde g + 1    }{\tilde z} \: \Ai' (\tilde z )
	\right] \,,\\
	\int \! \ud \theta 	\int \!\ud^2 \vec r_\perp \: e^{i\theta \frac{-k. \langle\pi_{-p}\rangle}{\kappa.q}}	\UP_2 &  \simeq
	4\pi^2 m^2 b_k \frac{1-s}{s}  \left [ \Ai_1(\tilde z)
	+   \frac{2 \tilde g - 1}{\tilde z} \: \Ai'(\tilde z)
	 \right]  \,,\\
	\int \! \ud \theta 	\int \!\ud^2 \vec r_\perp \: e^{i\theta \frac{-k. \langle\pi_{-p}\rangle}{\kappa.q}} \PP_1 
	& \simeq    4\pi^2  m^2 b_k \frac{1-s}{s} \:  \frac{\Ai(\tilde z)}{\sqrt{\tilde z}} \:   2\left( \tilde g - 1 + \frac{1}{s} \right)  \shd \,, \\
	\int \! \ud \theta 	\int \!\ud^2 \vec r_\perp \: e^{i\theta \frac{-k. \langle\pi_{-p}\rangle}{\kappa.q}}  \PP_2 
	& \simeq 4\pi^2 m^2 b_k  \frac{1-s}{s} \:  \frac{\Ai(\tilde z)}{\sqrt{\tilde z}} \: 2 (1 - \tilde g )  \: \shd \,,\\
	\int \! \ud \theta 	\int \!\ud^2 \vec r_\perp \: e^{i\theta \frac{-k. \langle\pi_{-p}\rangle}{\kappa.q}}  \EP_1 
	&  \simeq   4\pi^2  m^2 b_k \frac{1-s}{s} \:  \frac{\Ai(\tilde z)}{\sqrt{\tilde z}} \:   2 \left( \tilde g - 1 + \frac{1}{s} \right)  \shd \,, \\
	\int \! \ud \theta 	\int \!\ud^2 \vec r_\perp \: e^{i\theta \frac{-k. \langle\pi_{-p}\rangle}{\kappa.q}}  \EP_2
	& \simeq 4\pi^2 m^2 b_k  \frac{1-s}{s}  \:  \frac{\Ai(\tilde z)}{\sqrt{\tilde z}} \: 2(\tilde g - 1 ) \: \shd \,,\\
	\int \! \ud \theta 	\int \!\ud^2 \vec r_\perp \: e^{i\theta \frac{-k. \langle\pi_{-p}\rangle}{\kappa.q}} \PC_1 & \simeq 4\pi^2 m^2 b_k \frac{1-s}{s}   
	\left[ 
	 (2-\tilde g) \: \Ai_1(\tilde z) 
	 + \frac{2 +  \tilde g }{\tilde z} \: \Ai'(\tilde z) 
	\right] \,,\\
	\int \! \ud \theta 	\int \!\ud^2 \vec r_\perp \: e^{i\theta \frac{-k. \langle\pi_{-p}\rangle}{\kappa.q}} \PC_2 & \simeq  4\pi^2 m^2 b_k \frac{1-s}{s} 
	\left[ \tilde g \Ai_1(\tilde z)
	+  \frac{	2- \tilde  g   }{\tilde z}  \: \Ai' (\tilde z)
	\right] \,.
	\end{align}
	Combining these traces by plugging them into
	\begin{align}
	 \label{eq:rate.pp.eq}
    \frac{\ud \mathbb R_{\mathrm{NBW},j}}{\ud s} (  \sigma_p , \sigma_{q} )		
    &= \frac{\alpha}{16\pi^2 m^2 b_k^2 } \frac{s}{1-s} \:
    \int \! \ud \theta \int \! \ud^2 \vec r_\perp
    \: e^{i\theta \frac{-k. \langle\pi_{-p}\rangle}{\kappa.q}}
    \left[  \UP_j + \sigma_q \EP_j + \sigma_{p} \PP_j +  \sigma_p \sigma_{q}  \PC_j \right]\,,
	\end{align}
	we get the LCFA expressions for the decay rate per unit laser phase of a polarised photon in a polarisation state $\Lambda_j$, $j=1,2$, into a polarised electron-positron pair:
    \begin{align} \label{eq:rate.pp.final}
    \frac{\ud \mathbb R_{\mathrm{NBW},1}}{\ud s} (  \sigma_p , \sigma_{q} )	
     &= \frac{\alpha}{4b_k} 
     	\left[
     		\{(1 + \sigma_p \sigma_q (2-\tilde g) \} \Ai_1(\tilde z) \vphantom{\frac{1}{1}} 
     		-2(\sigma_p+\sigma_q) \left( 1 -\tilde g - \frac{1}{s} \right) \frac{\Ai(\tilde z)}{\sqrt{\tilde z}} \: \shd
     		\right. \nonumber \\
    & \qquad \quad + \left.
    \{ 2\tilde g+ 1 + \sigma_p \sigma_q (2 +\tilde g) \} \frac{\Ai'(\tilde z)}{\tilde z}
      	\right] \,, \\
    \frac{\ud \mathbb R_{\mathrm{NBW},2}}{\ud s} (  \sigma_p , \sigma_{q} )	
    &= \frac{\alpha}{4b_k}
    \left[
    \{(1 + \sigma_p \sigma_q \tilde g \} \Ai_1(\tilde z) \vphantom{\frac{1}{1}} 
    +2(\sigma_p-\sigma_q) \left( 1 -\tilde g \right) \frac{\Ai(\tilde z)}{\sqrt{\tilde z}} \: \shd
    \right. \nonumber \\
    & \qquad \quad + \left.
    \{ 2\tilde g - 1 + \sigma_p \sigma_q (2 -\tilde g) \} \frac{\Ai'(\tilde z)}{\tilde z}
    \right]
    \end{align}
    By introducing again the Stokes parameter for the incoming photon we arrive at Eq.~\eqref{eq:rate.pp.final.stokes} of the main text.

\end{widetext}

\bibliography{spin_pol}

\end{document}